\documentclass[epjST]{svjour}
\usepackage{graphics}
\usepackage{dirtytalk}
\graphicspath{ {./figures/} }
\usepackage{subfig}
\usepackage{xcolor}
\usepackage{amssymb}
\usepackage{amsmath}
\usepackage{multirow}
\usepackage[utf8]{inputenc}
\usepackage{amsmath}
\usepackage[demo]{graphicx}
\usepackage{caption}
\usepackage{graphics}
\usepackage{dirtytalk}
\begin{document}
%
\title{Is directed percolation class for synchronization transition robust with multi-site interactions?}
\author{Manoj C. Warambhe\inst{1,2}
\and Prashant M. Gade\inst{2}}

\institute{Department of CSE, GH Raisoni Skill Tech University, Nagpur, 
India\and Department of Physics,
RTM Nagpur University, Nagpur, India}
\abstract{
Coupled map lattice with pairwise local interactions 
is a well-studied system. However, in several situations, such as neuronal or social networks, multi-site interactions are possible. 
In this work, we study the coupled Gauss map in one dimension with 2-site, 3-site, 4-site and 5-site interaction. This coupling cannot be decomposed in pairwise interactions.
We coarse-grain the variable values by labeling the sites above 
$x^{\star}$ as up spin (+1) and the rest as down spin (-1) 
where $x^{\star}$ is the fixed point. We define flip rate $F(t)$ 
as the fraction of sites $i$ such that $s_{i}(t-1) \neq s_{i}(t)$ and 
 persistence $P(t)$ as the fraction of sites $i$ such that 
$s_{i}(t')=s_{i}(0)$ for all $t' \le t$. 
The dynamic phase transitions to a synchronized state
is studied above quantifiers. 
For 3 and 5 sites interaction, we find that at the critical point, $F(t) \sim t^{-\delta}$ with $\delta=0.159$ and $P(t) \sim t^{-\theta}$ with $\theta=1.5$. They match the directed percolation (DP) class. 
Finite-size and off-critical scaling is
 consistent with DP class.
For 2 and 4 site interactions, the exponent $\delta$ and behavior of $P(t)$ at critical point changes.  
Furthermore, we observe logarithmic oscillations over and above 
power-law decay at the critical point for 4-site coupling. Thus multi-site interactions can lead to
new universality class(es). 
}

%
\maketitle
\section{Introduction}
\label{intro}
A complex network has many interacting components that show the
various range of dynamical
behaviour\cite{andreev2019chimera,costa2011analyzing}.
Such a system is applicable in many real-world systems, including the
spreading of diseases in the populations \cite{cliff2019network}, complex 
biological pathways \cite{bhalla1999emergent}, social networks 
\cite{kanawati2015multiplex}, neuronal networks
\cite{andreev2019chimera,mishra2018dragon} etc.
Dynamical systems on such networks could be coupled
differential equations, coupled maps (CML), or cellular automata with a
decreasing degree of complexity. Most of these
models involve pairwise interactions between different elements.  
However, there are many systems where multi-site interaction is possible. 
This higher-order interaction is usually more complex and studied in many 
fields of science such as physics \cite{estrada2006subgraph}, mathematics 
\cite{chodrow2020configuration}, and computer science 
\cite{karypis1997multilevel}.  For instance, higher-order interactions are useful for modeling the functional brain connections\cite{petri2014homological}.
In this work, we study coupled maps with higher-order interactions.
Map-based models have less computational cost and can
reveal qualitative features.  Such interactions can change the nature of dynamic 
phase transitions.  Higher-order Kuramoto model can lead to explosive
synchronization \cite{millan2020explosive}. 
This is a discontinuous transition.
The self-organized behavior of coupled-phase oscillators with many-body interaction is
studied in \cite{skardal2020higher}.  This
coupling reproduces the rapid synchronization in many valid biological
networks. It has hysteresis and bistability indicating first-order
transition.
Such multi-site couplings 
are often studied on simplicial geometries 
\cite{gambuzza2021stability,bhattacharya2021higher,salnikov2018simplicial,ghorbanchian2021higher}.  
There are several reports of first-order transition in 
systems with simplicial couplings\cite{millan2020explosive,chutani2021hysteresis,kachhvah2022first,matamalas2020abrupt,jalan2022multiple}.  CML of logistic maps or neuronal maps with simplicial coupling is studied in \cite{naval,perc1}.
The above systems, many of which have real-world applications, motivate us to study the higher-order interaction
in a system of coupled chaotic maps.

In particular, we study the transition to synchronization.
We study CML with multi-site interaction. 
In this model, we study coupled map lattice
with 2-site, 3-site, 4-site, and 5-site interaction.
We retain the one-dimensional nature of underlying topology so that it is possible to compare it with one-dimensional transitions. 
Firstly, we find that the nature of the transition can be continuous
and not necessarily explosive. Standard finite-size scaling is observed.
(In explosive synchronization, it is claimed that the finite-size scaling
is atypical\cite{d2019explosive}.) 
All exponents including relatively less universal persistence 
exponent match with expected values for one-dimensional DP for the 3-site and 5-site interaction. The situation changes for the 2-site or 4-site interaction.

\begin{figure}[h]%
	\centering
	\includegraphics[width=0.5\textwidth]{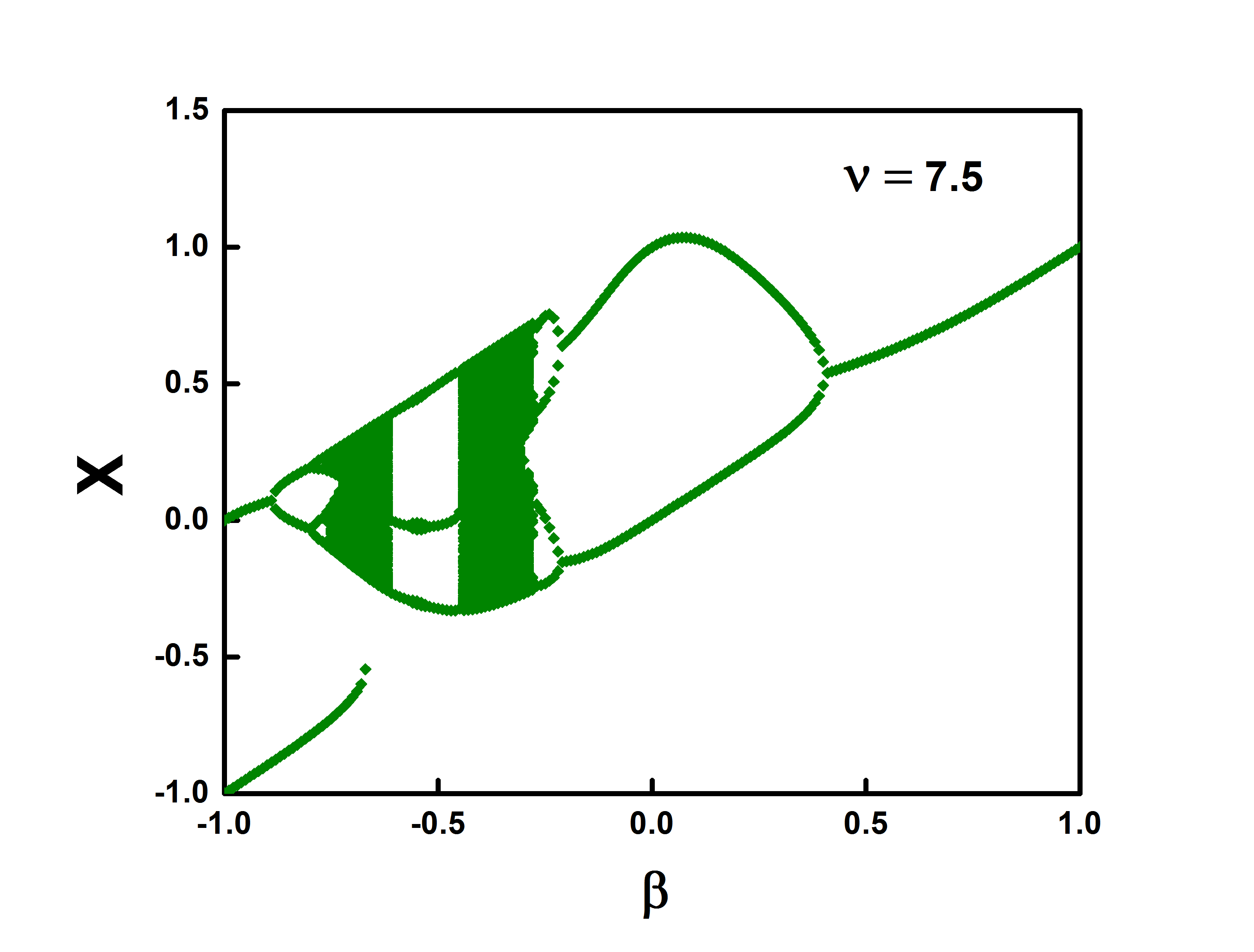}
	\caption{Bifurcation diagram for Gauss map for $\nu=7.5$. It shows the reverse
		period-doubling, period-adding as well as chaos. }
	\label{fig1}
\end{figure}

The Gauss map is a $1D$ map based on the Gaussian
exponential function \cite{patidar2006co}.
Unlike logistic, tent, or circle maps,
this is a map defined on the entire real line and not an interval.
Such maps are relatively less studied.
On changing the map parameter values, this map shows the diverse range
of dynamical behaviour including reverse period doubling,
period adding and chaos.
In standard form, it can be defined as:
\begin{equation}
	f(x)=\exp(-\nu x^2)+\beta, \hspace{1cm} x\in R
	\label{eq1}
\end{equation}
Where $\nu$ and $\beta$ are the control parameters of the given map. We plot the bifurcation diagram of the Gauss map for $\nu=7.5$ [see \textbf{Fig.\ref{fig1}}].
We plot all sites as a function of map parameter $\beta$. The bifurcation plot displays distinct dynamical behavior including the reverse period-doubling, period-adding
and chaos.

We study the CML with multi-site interaction in the Gauss map with positive 
coupling.  Our studies are the viewpoint of phase transition, and this phase 
transition is identified by the appropriate order parameter. We consider local 
persistence $P(t)$ and flip rate $F(t)$ as order parameters. Persistence is 
the probability that a nonequilibrium field value does not change the sign up to time $t$ during stochastic evolution \cite{majumdar1999persistence}. This 
is a non-Markovian quantity and experimental investigations are difficult for 
obvious reasons.  In several systems, persistence decays algebraically with 
time at the critical point as $P(t) \sim t^{-\theta}$ where $\theta$ is the decay exponent known as 
the persistence exponent.  We need the knowledge of entire time evolution to 
find this quantity.

For CML, transition to a frozen or partially frozen state in a coarse-grained 
sense can be characterized by persistence and power-law behaviour may be 
obtained only at the critical coupling. However, the persistence exponent is 
not universal. As noted below, several $1D$ systems in the DP universality 
class have persistence exponent $3/2$, and several systems in directed Ising universality have persistence exponent unity \cite{shambharkar2019universality}. 
The persistence exponent in the $1D$ DP model such as the
Domany-Kinzel model \cite{hinrichsen1998numerical}, Ziff-Gulari-Barshed model 
\cite{albano2001numerical}, site percolation \cite{fuchs2008local} and $1D$ 
circle map \cite{menon2003persistence,jabeen2005dynamic} is observed to be 
$3/2$. (We note that the synchronization transition between two coupled CML is found to be in bounded KPZ universality class for smooth maps\cite{ahlers}. But, we focus on a single CML in this work.) The transition to a random pattern is frozen in the coarse-grained
state (not synchronized) can be in DP universality
class in CML \cite{pakhare2020novel,pratik1} and shows persistence exponent close to
$3/2$. Although not universal, we can observe the same persistence exponent in a significant set of models belonging to the same universality class. Hence, it is important. In the above cases of coupled maps, flip rate $F(t)$ is studied as
a measure of lattice activity at a given time $t$. Recently,
a new universality class pertaining to transition to a period
$n$ absorbing state has been observed in coupled map lattice
and has been investigated using the flip rate as a quantifier\cite{pratik1}. (It is also modeled by the cellular automata model in \cite{divya1}.) 
We use the similar measure here. We denote sites with a variable value greater than a fixed point as spin (+1) and those with a value less than a fixed point
as spin (-1).  For the case of period-3 synchronization,
flip rate is defined as the fraction of sites that change their spin state from state at time $t-3$\cite{pratik1}. If the underlying coarse-grained periodicity is 2, it is the fraction of sites that changed its state at time $t$ from the state at $t-2$\cite{naval,pakhare2020novel}. Thus, any departure from the expected underlying dynamics is the flip rate. In our case, the flip rate $F(t)$ is the fraction of sites that change their 
spin state at  $t$ from the previous state at  $(t-1)$.

DP is the most widely observed universality class in the nonequilibrium 
system from the viewpoint of phase transition from the active state
to the absorbing state.  It is observed in several systems including
transition to turbulence \cite{hof2023directed} or transition between two 
turbulent states of electroconvection in nematic liquid crystals
\cite{takeuchi2009experimental}. Janssen–Grassberger conjecture 
\cite{janssen1981nonequilibrium,grassberger1981phase}
stated the conditions for DP transition as \cite{grassberger1981phase} 
“the universality class of DP contains all continuous transitions from a
dead or absorbing state to an active one with a single scalar order parameter, 
provided the dead state is not degenerate and provided some technical points 
are fulfilled: short-range interactions both in space and time, the 
nonvanishing probability for any active state to die locally, translational 
invariance (absence of frozen randomness), and absence of multicritical 
points”. The question is whether all these conditions are necessary and what 
happens if we relax some or all of them.  Relaxation in some of these conditions  still leads to the directed 
percolation universality class\cite{bhoyar2022robustness}. Most of the models studied have pairwise interactions. We study interaction
that cannot be decomposed as pairwise interaction and
check the universality class of absorbing state transition in this system.

For CML with pairwise interaction, there are numerous instances of dynamic 
transitions in the DP universality class. For the coupled Gauss map, it has been 
observed that the transition to a frozen state in a coarse-grained sense
is in DP universality class \cite{pakhare2020novel}. Chate and Mannevile 
studied the coupled piecewise linear discontinuous map and showed that the
transition led to DP class behavior \cite{chate1988spatio}.
It has been also shown that
unidirectional coupling and asymmetric coupling are
not relevant perturbations for
the DP class \cite{tretyakov1997phase}.
In \cite{bhoyar2022robustness}, the robustness of the DP class
in the presence of
recovery time, memory, external forcing, or quenched disorder
was checked. It was found that these
perturbations do not alter the universality class of the
transition. In this work, we find that the DP class is robust for
 coupled Gauss maps for 3-site and 5-site interaction. Thus some multi-site interactions are not a relevant perturbation to the DP universality class. However,
for 2-site or 4-site interaction, we obtain new exponents, and the class changes.

\section{The Model }
We consider the following model. Consider a lattice of length $N$. We
denote the local
variable at site $i$ as  $x_{i}(t)$. The initial
conditions are chosen as a uniform random number
in the interval $[0, 1]$.
We use
the periodic boundary conditions.
The evolution proceeds in a synchronous manner. The time
evolution of $x_{i}(t)$ at discrete time $t$ for 2-sites, 3-site, 4-site and 5-site interaction are given by
\begin{equation}
	x_{i}[t+1]=(1-\epsilon) f[x_{i}(t)]+\epsilon f[x_{i-1}(t)]
	f[x_{i+1}(t)]
	\label{eq2}
\end{equation}
\begin{equation}
	x_{i}[t+1]=(1-\epsilon) f[x_{i}(t)]+\epsilon f[x_{i}(t)] f[x_{i-1}(t)]
	f[x_{i+1}(t)] 
	\label{eq3}
\end{equation}
\begin{equation}
	x_{i}[t+1]=(1-\epsilon) f[x_{i}(t)]+\epsilon f[x_{i-1}(t)] 
	f[x_{i+1}(t)] f[x_{i-2}(t)] f[x_{i+2}(t)]
	\label{eq4}
\end{equation}
\begin{equation}
	x_{i}[t+1]=(1-\epsilon) f[x_{i}(t)]+\epsilon f[x_{i}(t)] f[x_{i-1}(t)] 
	f[x_{i+1}(t)] f[x_{i-2}(t)] f[x_{i+2}(t)]
	\label{eq5}
\end{equation}
where $t$ is time index and $i=1,2,3,...,N$ is the site index.
The difference from the usual CML is that this coupling cannot be decomposed in pairwise interactions.
In the above equation, $\epsilon$ is
the coupling parameter, which is a measure of the strength
of coupling between site $i$ and its neighbours.
At $\epsilon=0$ we recover the function
$f(x)$ and all sites evolve independently of
each other. The function $f(x)$ is the Gauss map which is
defined in \textbf{Eq. \ref{eq1}}.
Let, $x^{\star}$ be the non-zero fixed point of the Gauss map which is
obtained using the bisection method.
We associate spin is $s_i(t)$ with
site $i$ at time $t$.  If the variable value $x_i(t)> x_{i}^{\star}$, $s_i(t)=1$ and
if $x_i(t)< x_{i}^{\star}$, $s_i(t)=-1$.
We define two quantifiers local persistence $P(t)$ and flip rate $F(t)$ as follows:\\

$\textbf{Persistence:}$
\textit{The fraction of sites that did not
	alter their spin value even once throughout all time steps are
	referred to as persistent sites. The
	fraction of persistent sites denoted by $P(t)$ is 
	persistence at time $t$.
	Thus, the persistence $P(t)$ at time $t$ is  fraction of sites $i$ such
	that $s_{i}(t')=s_{i}(0)$ for all $t' \leq t$.} \\

$\textbf{Flip rate:}$
\textit{Flip rate $F(t)$ is the fraction of sites at time $t$ that
	changed their spin state  from the state at
	time $(t-1)$.
	In other words,
	flip rate $F(t)$ is the fraction of sites $i$ such that
	$s_{i}(t-1) \neq s_{i}(t)$.}

\section{Results and Discussion}
\subsection{Coupled map lattice with 5-site interaction}
\begin{figure}[h]%
	\centering
	\includegraphics[width=0.65\textwidth]{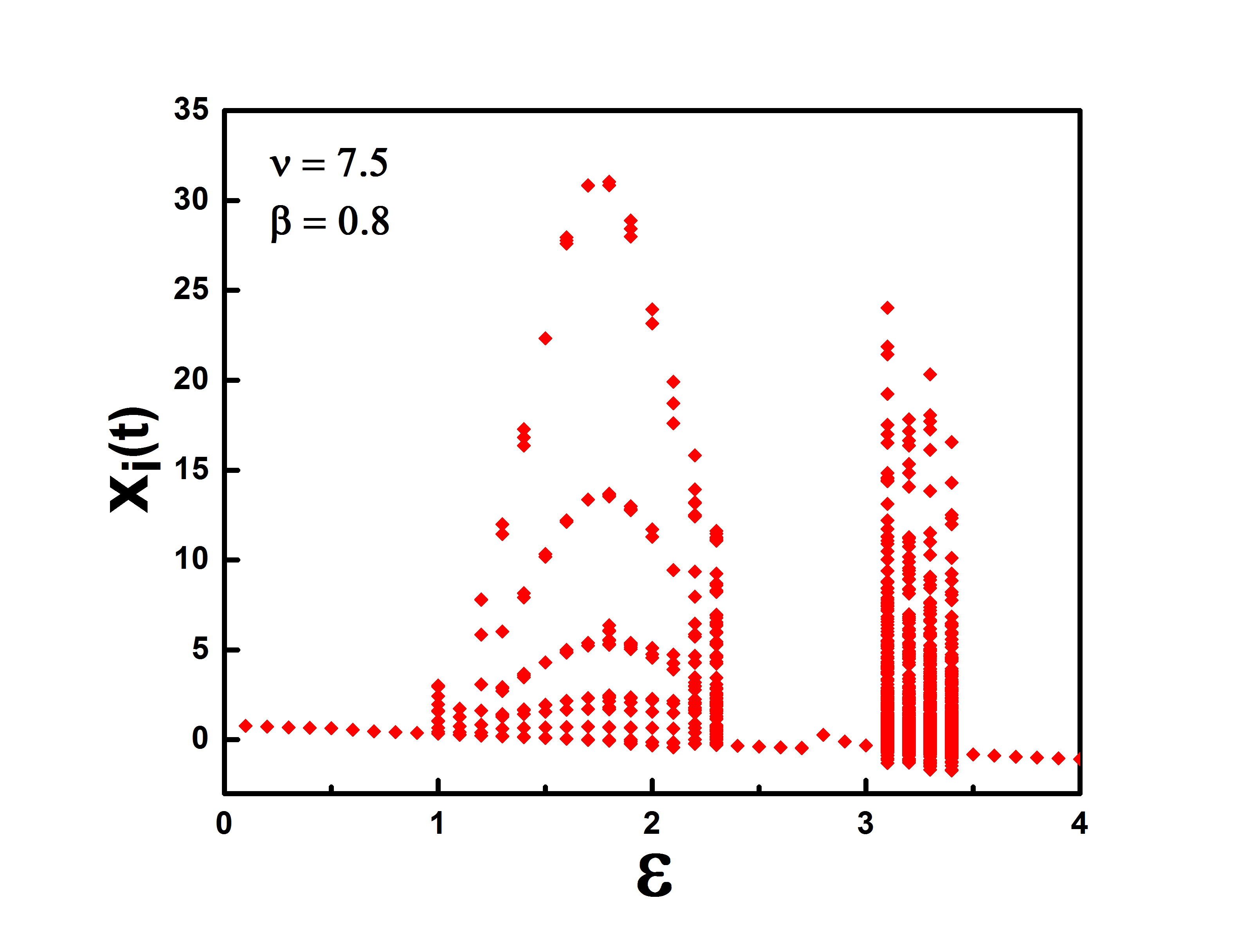}
	\caption{Plot of bifurcation diagram for 5-site interaction coupled Gauss maps
		where $x_i(t)$ is plotted as a function of
		$\epsilon$. We fix $\nu=7.5$ and $\beta=0.8$ and wait for $t=10^5$ time steps.}
	\label{fig2}
\end{figure}

First, we study our model for a 5-site interaction coupled Gauss map. The bifurcation diagram is the most simple method for displaying dynamic behavior while varying a control parameter. We plot the bifurcation diagram for the 5-site coupled Gauss map for map parameters $\nu=7.5$ and $\beta=0.8$ [see \textbf{Fig.\ref{fig2}}]. We consider
$N=500$ and wait for $t=10^5$. The bifurcation
diagram displays the all values of sites $x_i(t)$ as a function of coupling parameter $\epsilon$ at $t=10^5$. The bifurcation plot shows a clear synchronization transition for a larger value of $\epsilon$, which we investigate in detail.

\begin{figure}[hbt!]
	\centering
	\scalebox{0.235}{
		\includegraphics{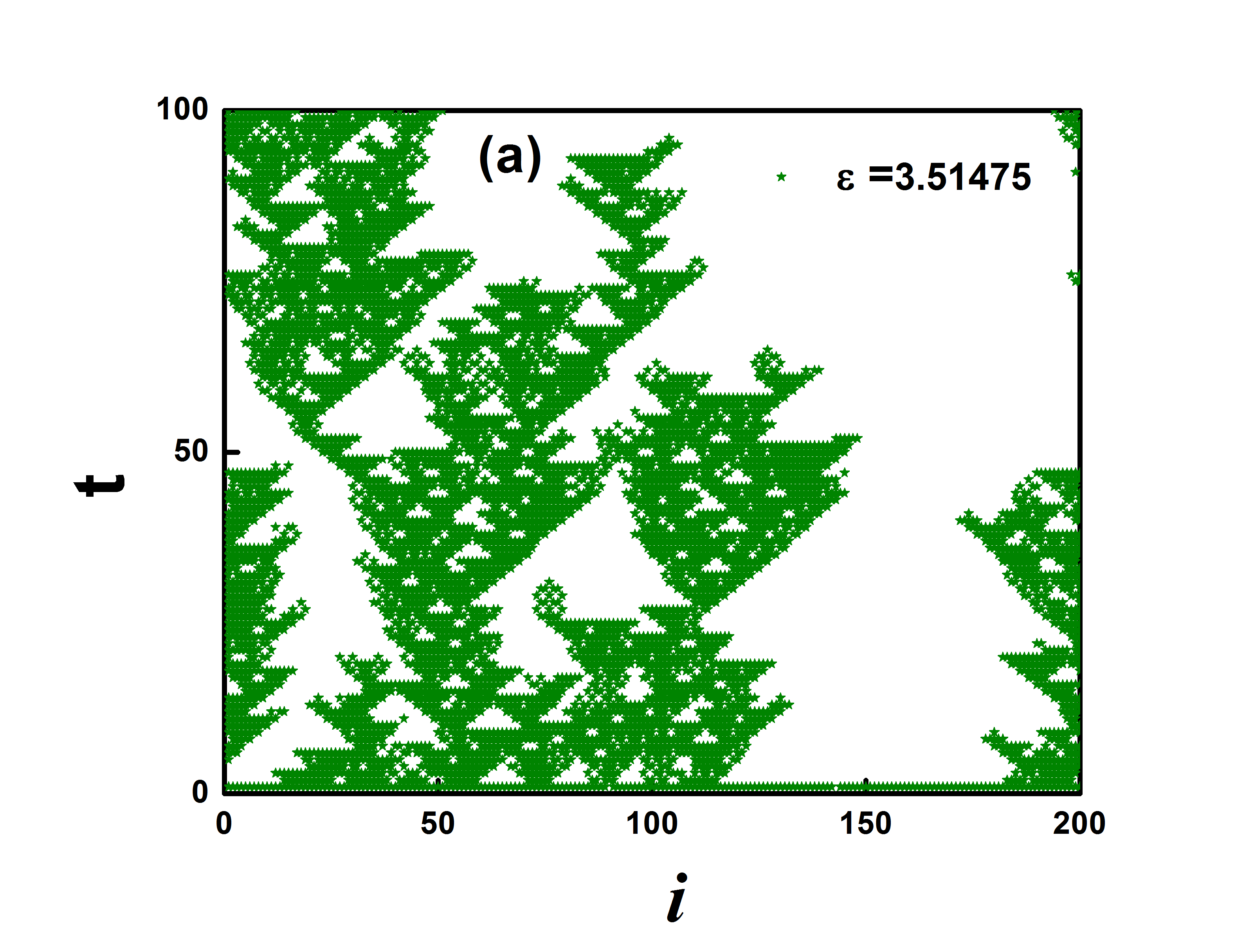}}
	\scalebox{0.235}{
		\includegraphics{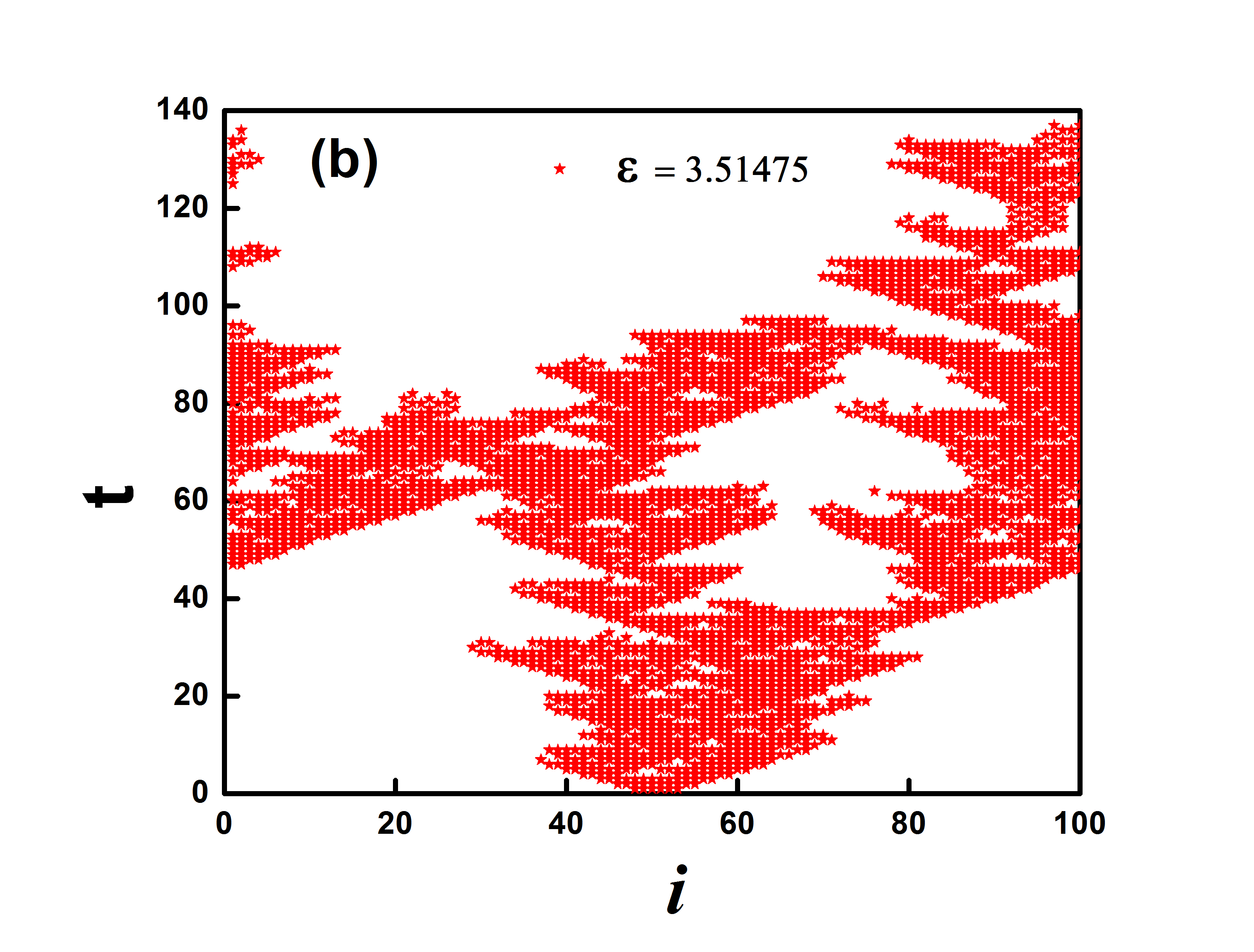}}
	\caption{ \textbf{(a)} Space-time plot of defects for
		5-site coupled Gauss map 
		for $N=100$ and $\epsilon=3.51475$ for random initial conditions.
		\textbf{(b)} Space-time plot of defects for coupled Gauss map for $N=100$ and $\epsilon=3.51475$.
		At $t=0$ $x(i,0)=x^*$ for all $i$ except three sites in the center. }
	\label{fig3}
\end{figure}

We also study the space-time diagram for visualizing the defects for the 5-site interaction coupled Gauss map. We plot it for discrete time $t$ as a function of site index $i$ (where $i=1,..., N$) for $N=100$ and $\epsilon=3.51475$. The behavior of such defects is plotted in \textbf{Fig. \ref{fig3} (a)} and in \textbf{Fig. \ref{fig3} (b)}. (Defects are the sites for which
$\vert x_i(t)-x^*\vert >\eta$ where $\eta=0.1$.) We observe the spatiotemporal intermittency (STI) in two different regions for both plots. There is no \textit{`solitonlike'} structure observed in this regime. A similar evolution is observed in \cite{janaki2003evidence} for coupled map lattice defined in a standard manner (see Figure 1 and Figure 2 of \cite{janaki2003evidence} ).

Saturation of persistence usually implies that the activity
has stopped fully or partially. Hence the synchronization
is accompanied by saturation of persistence. Thus, the order
parameter becoming zero is accompanied by saturation
of persistence. On the other hand, the active state will lead
to persistence becoming zero if activity continues at all
lattice sites.
Thus, there is a critical value of coupling $\epsilon_c$ such that
for $\epsilon>\epsilon_c$ the order parameter goes to zero and
persistence saturates while for $\epsilon<\epsilon_c$, the order
parameter saturates and persistence goes to zero.
For continuous transition, we observe power-law decay of both persistence and
order parameter at the critical point $\epsilon_c$.
At such a critical point,
the  persistence and order parameter algebraically decays as:
\begin{equation}
	P(t) \sim t^{-\theta}
\end{equation}
\begin{equation}
	O(t) \sim t^{-\delta}
\end{equation}
The exponent $\theta$ is known as the persistence exponent and $\delta$ is
the order parameter exponent. The exponents $\theta$ are not universal and depend
on the details of the evolution of the system.

\begin{figure}[hbt!]
	\centering
	\scalebox{0.235}{
		\includegraphics{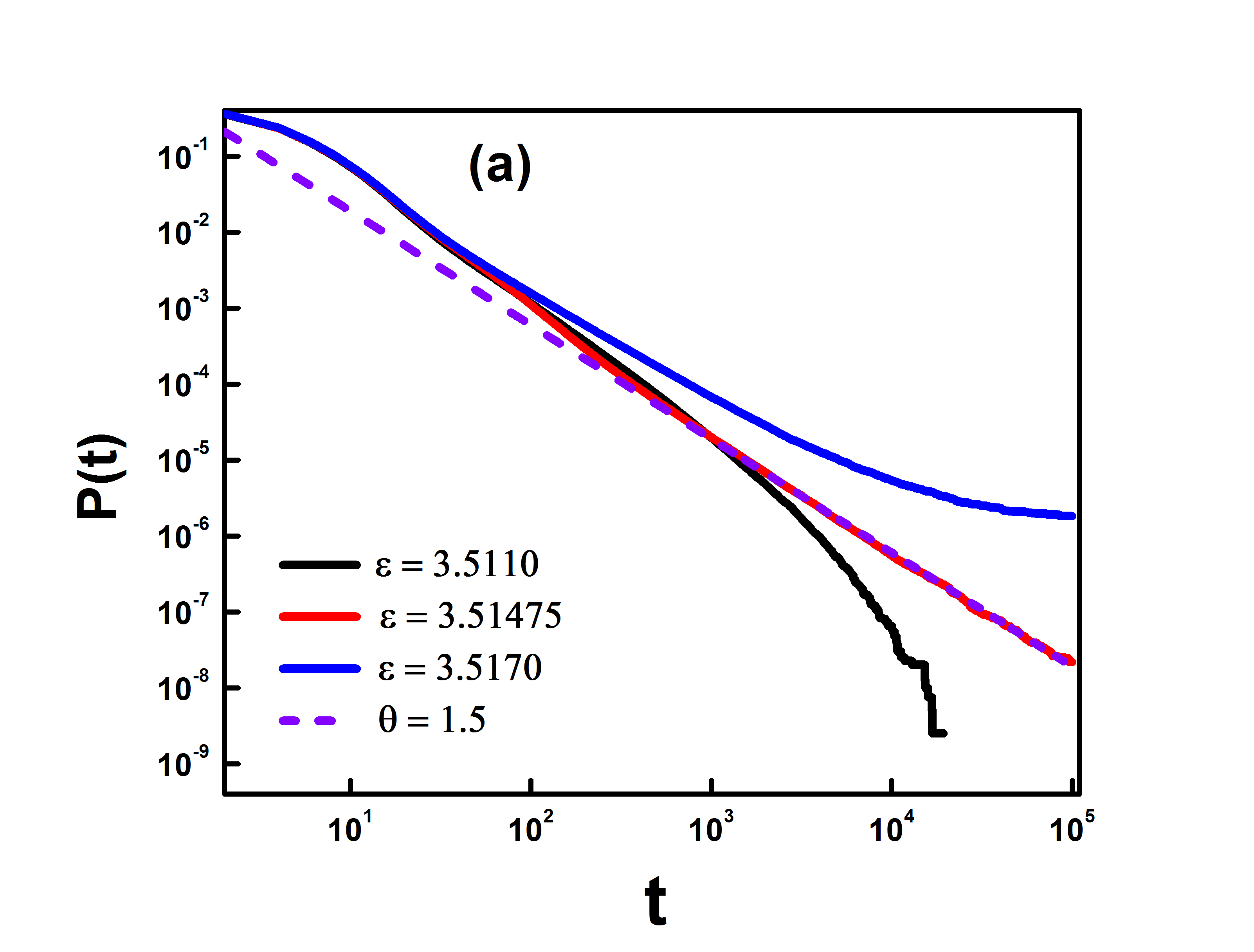}}
	\scalebox{0.235}{
		\includegraphics{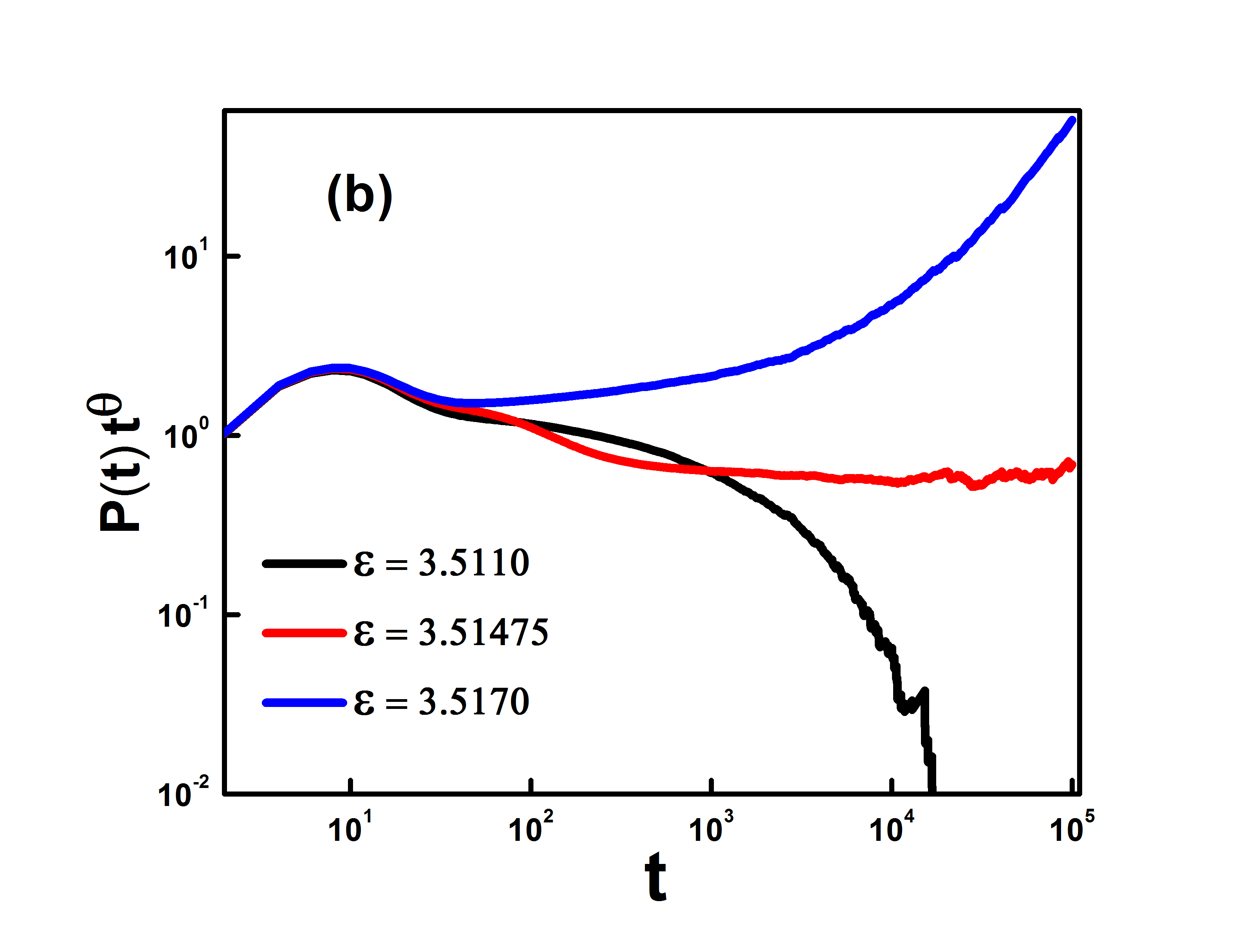}}
	\caption{ \textbf{(a)} Plot of persistence $P(t)$ as a function of time $t$ for
		$\epsilon < \epsilon_c$, $\epsilon > \epsilon_c$ and at $\epsilon=\epsilon_c$ for $N=2\times 10^5$. We average over a $1.2 \times 10^3$ configuration.
		We observe that $P(t) \sim t^{-\theta}$, $\theta=1.5$ at $\epsilon=\epsilon_c=3.51475$.
		\textbf{(b)} Plot of $P(t)t^{\theta}$ as a function of $t$, for $\epsilon < \epsilon_c$,
		$\epsilon > \epsilon_c$ and at $\epsilon=\epsilon_c$ for the same data as \textbf{(a)}.}
	\label{fig4}
\end{figure}

\begin{figure}[hbt!]
	\centering
	\scalebox{0.235}{
		\includegraphics{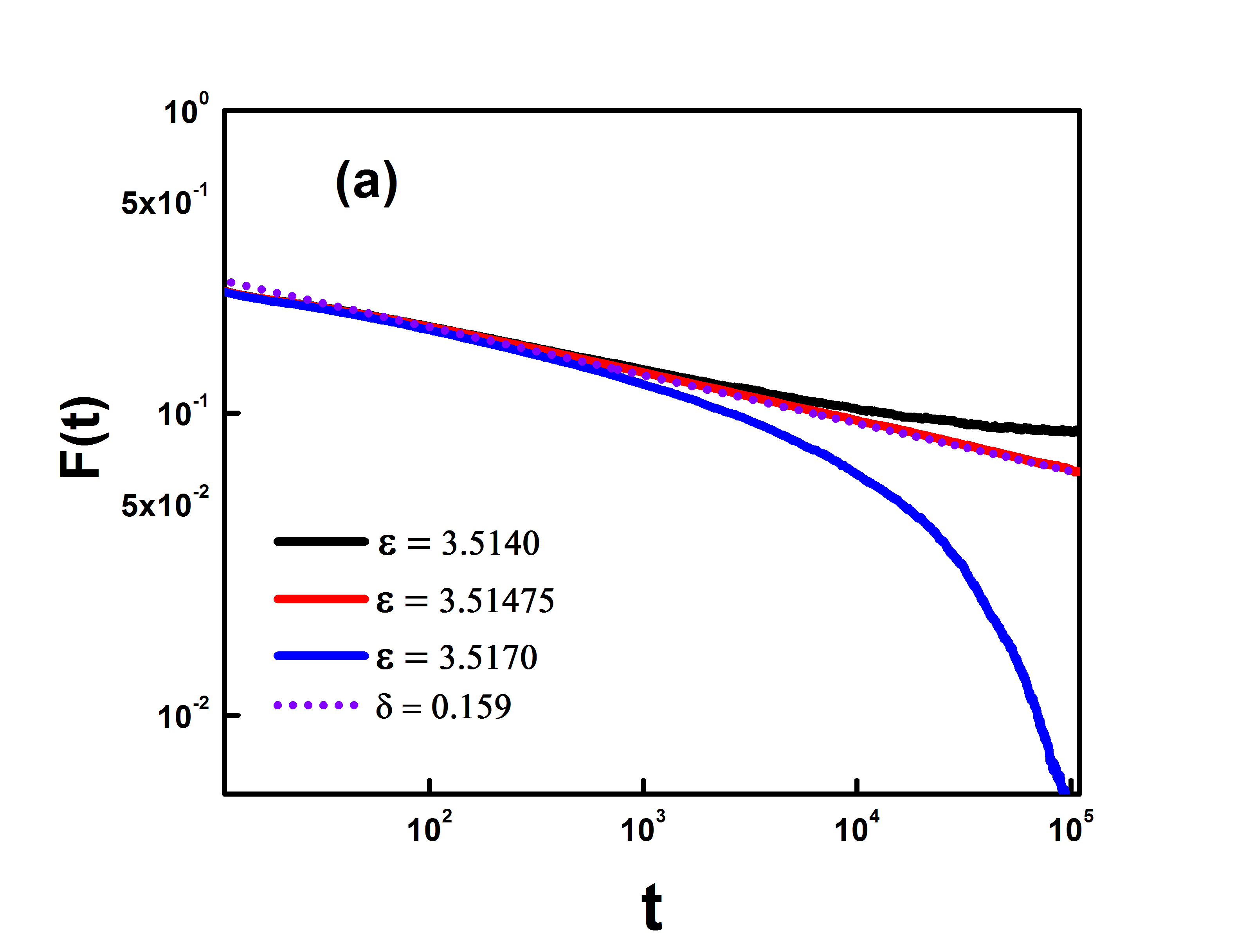}}
	\scalebox{0.235}{
		\includegraphics{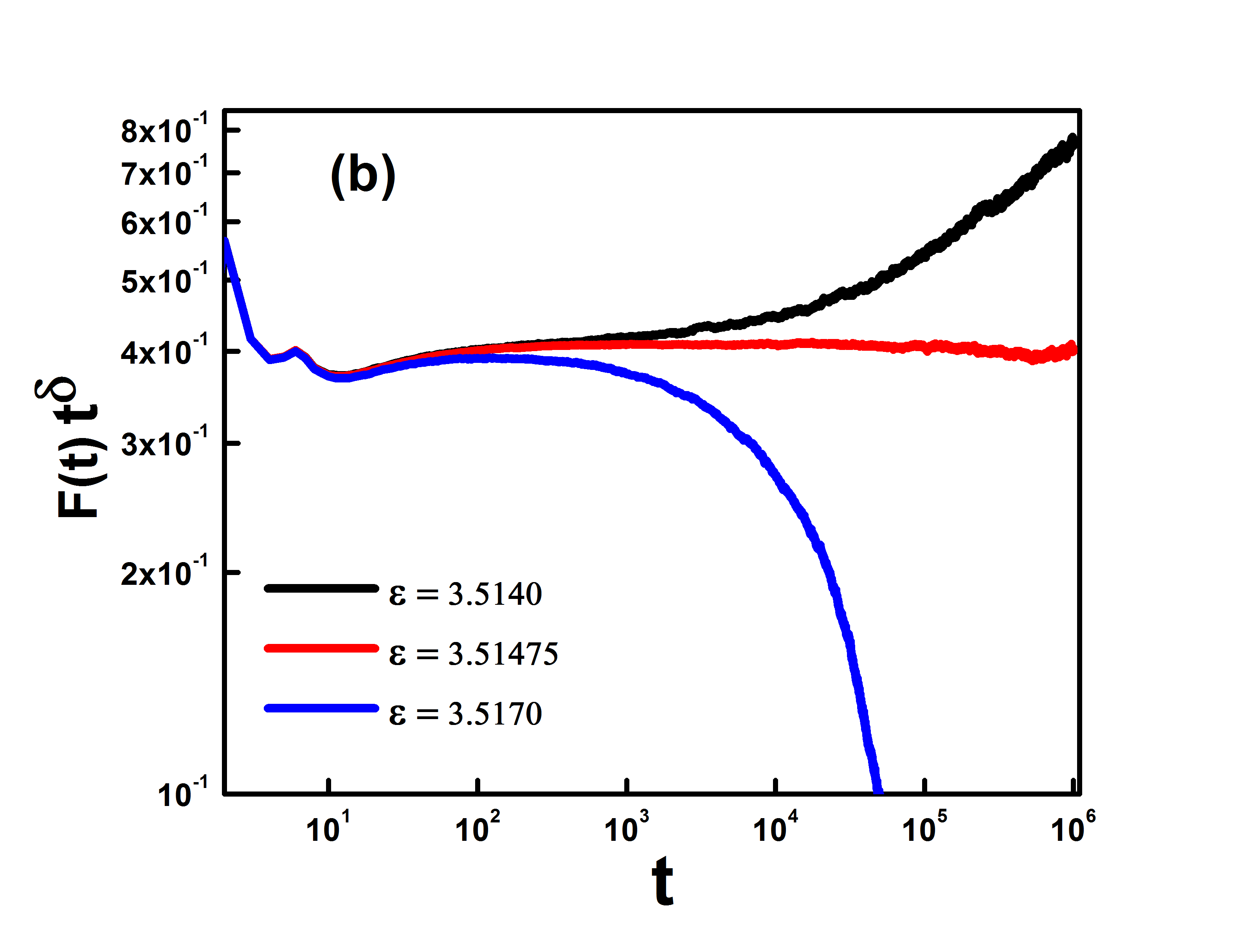}}
	\caption{ \textbf{(a)} Plot of Flip rate $F(t)$ as a function of time $t$ for
		$\epsilon < \epsilon_c$, $\epsilon > \epsilon_c$ and at $\epsilon=\epsilon_c$ for $N= 2\times 10^5$. 
		We average over a $1.2 \times 10^3$ configuration.
		We observe $F(t) \sim t^{-\delta}$, $\delta=0.159$ at $\epsilon=\epsilon_c=3.51475$.
		\textbf{(b)} Plot of $F(t)t^{\delta}$ as a function of time $t$, for $\epsilon < \epsilon_c$, $\epsilon > \epsilon_c$ and at $\epsilon=\epsilon_c$ for the same data as \textbf{(a)}.}
	\label{fig5}
\end{figure}

We simulate our 5-site interaction CML Gauss map for $\beta=0.8$, $\nu=7.5$, and
$N=2\times 10^5$. The initial conditions are random and
we average over $1.2\times 10^3$ configurations. We plot
both $P(t)$ and $F(t)$ as a function of
time $t$ [see \textbf{Fig. \ref{fig4} (a)} and \textbf{Fig. \ref{fig5} (a)}].
From these plots, it can be seen that for $\epsilon < \epsilon_c$,
$F(t)$ saturates and the $P(t)$ decays exponentially as expected.
Similarly,
for $\epsilon > \epsilon_c$,  $P(t)$ saturates and  $F(t)$ decays
exponentially. At critical point $\epsilon=\epsilon_c=3.51475$, we observe
$P(t) \sim t^{-\theta}$ and $F(t) \sim t^{-\delta}$.
The persistence decays with the exponent
$\theta=1.5$ while flip rate decays with exponent $\delta=0.159$.
If $P(t) \sim t^{-\theta}$ and $F(t) \sim t^{-\delta}$ then
$P(t)t^{\theta}$ and $F(t)t^{\delta}$ should be constant
asymptotically. We
plot $P(t)t^{\theta}$ and $F(t)t^{\delta}$ as a function of
$t$ for $\epsilon < \epsilon_c$,
$\epsilon > \epsilon_c$ and at $\epsilon=\epsilon_c$
[see \textbf{Fig. \ref{fig4} (b)} and \textbf{Fig. \ref{fig5} (b)} ]. 
We find that both 
the quantity tends to a constant
for $\epsilon=\epsilon_c$ as $t\rightarrow \infty$. For $\epsilon\neq \epsilon_c$, the curve displays upward or downward curvature.
This is alternative evidence
for the confirmation of the obtained values of the exponent.
The exponent $\delta$ and $\theta$ match with DP exactly.


To confirm the nature of transition, we carry finite-size and off-critical scaling to compute the exponents $z$ and $\nu_{\parallel}$. This exercise is carried out for both order parameter $F(t)$ as well as persistence $P(t)$. To compute dynamic exponent  $z$, we carry out finite-size scaling simulations for distinct lattice sizes at $\epsilon=\epsilon_c$. To compute $\nu_{\parallel}$, we carry out off-critical scaling. For this, we simulate a very large size lattice for various values of $\epsilon$ both
above and below $\epsilon_c$. We consider a lattice of
size $N=10^5$ large enough so that finite size corrections
are not important. 

The scaling law for $P(t)$ and $F(t)$ hold to expect following relations:
\begin{equation}
	P_N(t)=t^{-\theta}G(t/N^z,t {\Delta}^{\nu_{\parallel}})
\end{equation}

\begin{equation}
	F_N(t)=t^{-\delta}G(t/N^z,t {\Delta}^{\nu_{\parallel}})
\end{equation}

where $F$ and $G$ are scaling function and
$\Delta=\mid \epsilon-\epsilon_c \mid$
departure from
critical point.

Thus for $\Delta=0$, {\it{i.e.}} $\epsilon=\epsilon_c$, we obtain
\begin{equation}
	P_N(t)=t^{-\theta}G(t/N^z)
\end{equation}
\begin{equation}
	F_N(t)=t^{-\delta}G(t/N^z)
\end{equation}

and as $N\rightarrow \infty$, we obtain
\begin{equation}
	P_N(t)=t^{-\theta}G(t {\Delta}^{\nu_{\parallel}})
\end{equation}
\begin{equation}
	F_N(t)=t^{-\delta}G(t {\Delta}^{\nu_{\parallel}})
\end{equation}

\begin{figure}[hbt!]
	\centering
	\scalebox{0.235}{
		\includegraphics{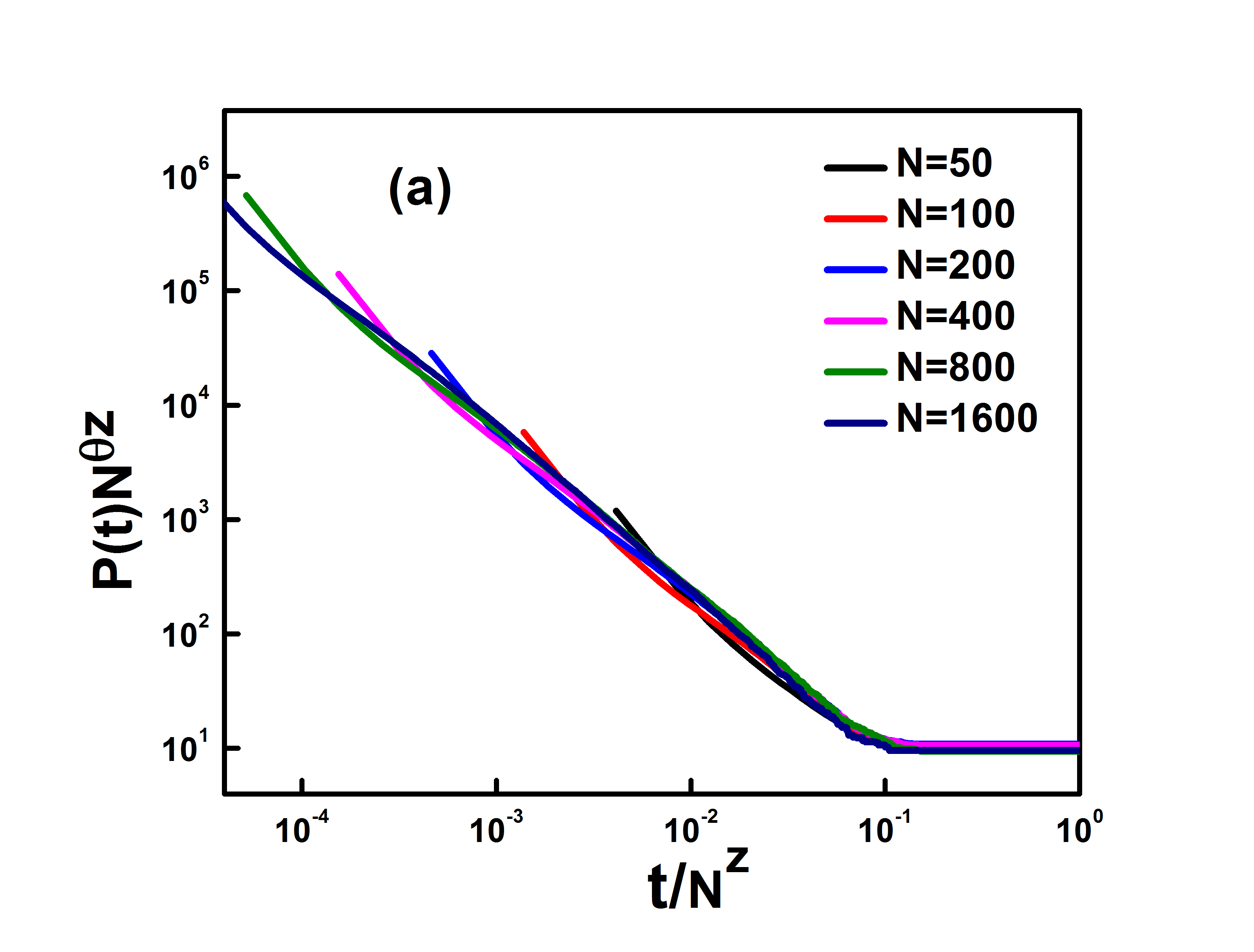}}
	\scalebox{0.235}{
		\includegraphics{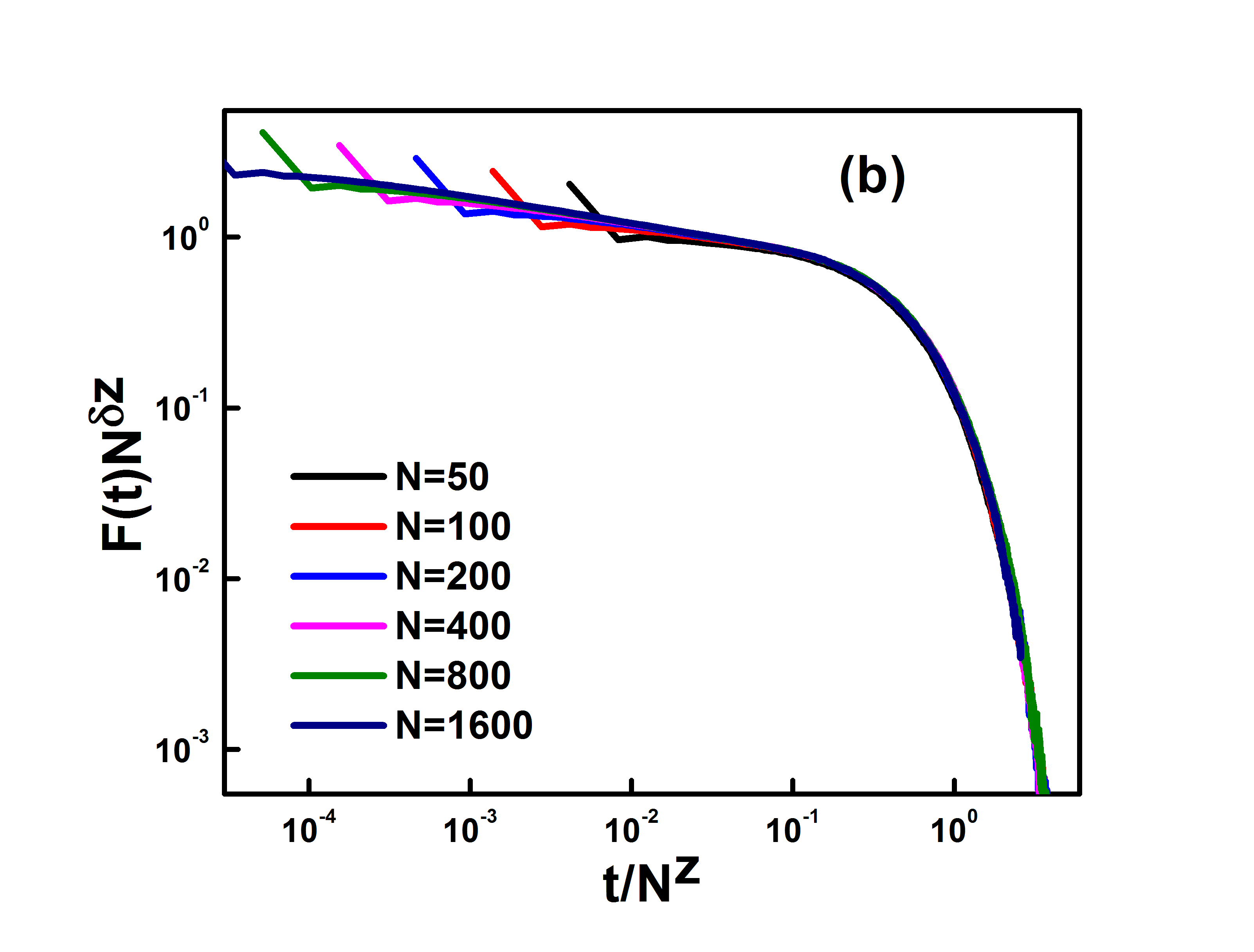}}
	\caption{\textbf{(a)} Plot of $P(t) N^{\theta z}$ as a function of
		$t/N^z$ at $\epsilon=\epsilon_c=3.51475$. 
		A good scaling collapse
		is observed at $z=1.58$ and $\delta=0.159$.
		\textbf{(b)} Plot lot $F(t) N^{\delta z}$ as a function of $t/N^z$ at $\epsilon=\epsilon_c=3.51475$.
		The fine collapse is observed at $z=1.58$ and $\theta=1.5$.}
	\label{fig6}
\end{figure}

We simulate the lattice for $N=50, 100, 200, 400, 800$, and $1600 $ and
average over
more than $1.2 \times 10^4$ configurations.
We plot $P(t) N^{\theta z}$ as a function of
$t/N^z$ at critical point $\epsilon=\epsilon_c=3.51475$
and the good scaling collapse
is obtained at $z=1.58$ which is shown in \textbf{Fig. \ref{fig6} (a).}
Similarly, we
also plot $F(t) N^{\delta z}$ as a function of $t/N^z$ and the fine collapse is
obtained at $z=1.58$ [see \textbf{Fig. \ref{fig6} (b)}].
The obtained values of dynamic exponent for both $P(t)$
and $F(t)$ exactly match the dynamic exponent $z=1.58$ of the DP class.

\begin{figure}[hbt!]
	\centering
	\scalebox{0.235}{
		\includegraphics{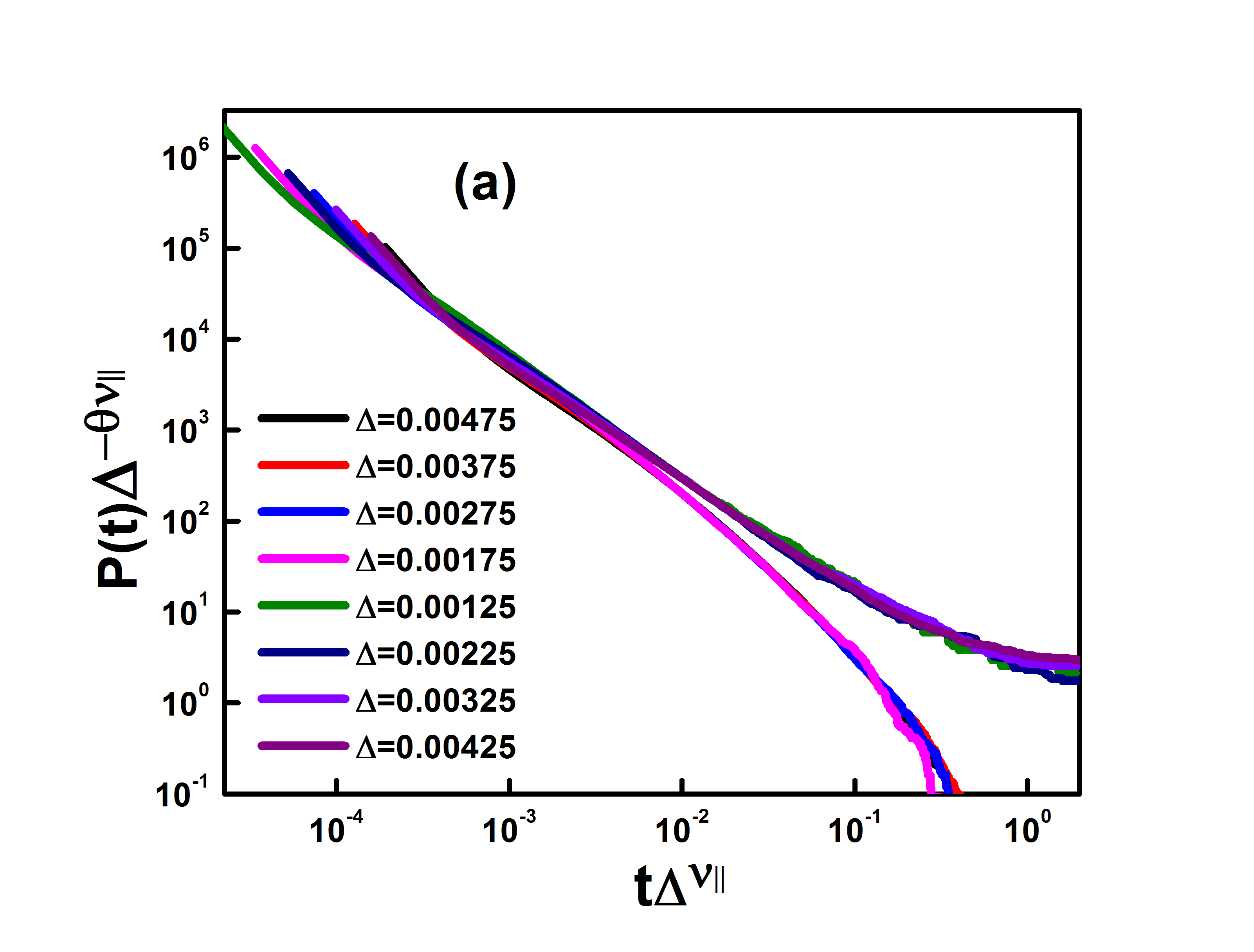}}
	\scalebox{0.235}{
		\includegraphics{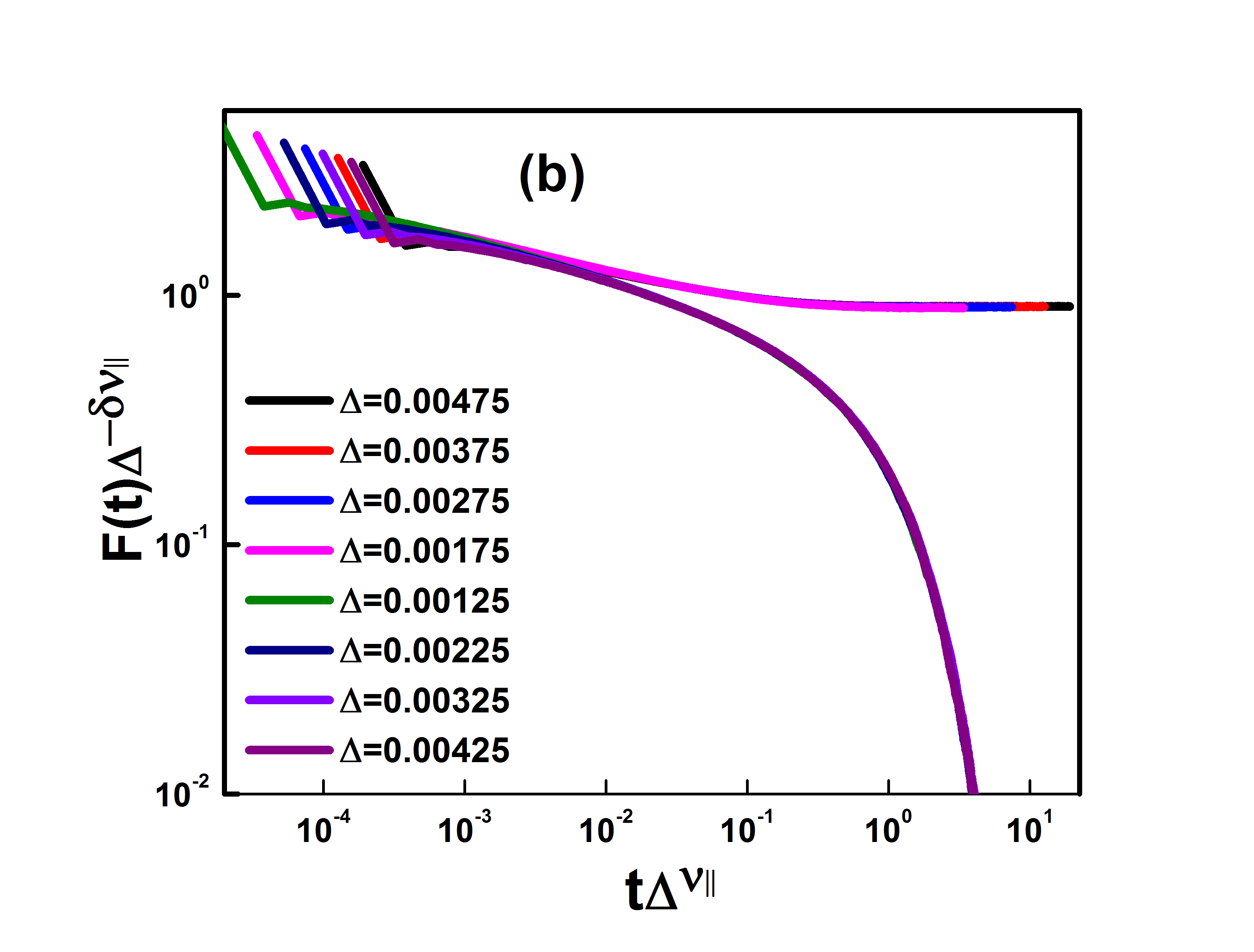}}
	\caption{ We simulate a lattice of size $N=1\times10^5$ and each 
		average over more than
		$800$ configurations for several values of $\epsilon$
		above and below $\epsilon_c$.
		\textbf{(a)}  We plot $P(t){\Delta}^{-{\theta}{\nu_{\parallel}}}$
		as a function of
		$t {\Delta}^{\nu_{\parallel}}$ for $\epsilon < \epsilon_c$
		and for	$\epsilon > \epsilon_c$.
		The good scaling collapse is observed at $\nu_{\parallel}=1.73$.
		\textbf{(b)} We plot $F(t) {\Delta}^{-{\delta}{\nu_{\parallel}}}$
		as a function of
		$t {\Delta}^{\nu_{\parallel}}$.
		The good scaling collapse is observed at $\nu_{\parallel}=1.73$.}
	\label{fig7}
\end{figure}

We note that for $\epsilon < \epsilon_c$,
$P(t)$ decays to zero exponentially and
$F(t)$ saturates.
For $\epsilon > \epsilon_c$, $F(t)$ decays to zero and $P(t)$
saturates. We simulate a large lattice of
size $10^5$ and average over more than 800 configurations
for several values of $\epsilon$ above and below $\epsilon_c$.
We plot $P(t){\Delta}^{-{\theta}{\nu_{\parallel}}}$
as a function of $t {\Delta}^{\nu_{\parallel}}$
and the good scaling collapse is obtained
for $\nu_{\parallel}=1.73$. This behavior is shown in\textbf{ Fig. \ref{fig7} (a)}.
Similarly, we plot $F(t){\Delta}^{-{\delta}{\nu_{\parallel}}}$
as a function of
$t {\Delta}^{\nu_{\parallel}}$ and a good scaling collapse is
obtained at $\nu_{\parallel}=1.73$.
This collapse is shown in \textbf{Fig. \ref{fig7} (b)}.
The obtained values of $\nu_{\parallel}=1.73$ match with
the exponent of the DP class exactly.

\begin{figure}[hbt!]
	\centering
	\scalebox{0.235}{
		\includegraphics{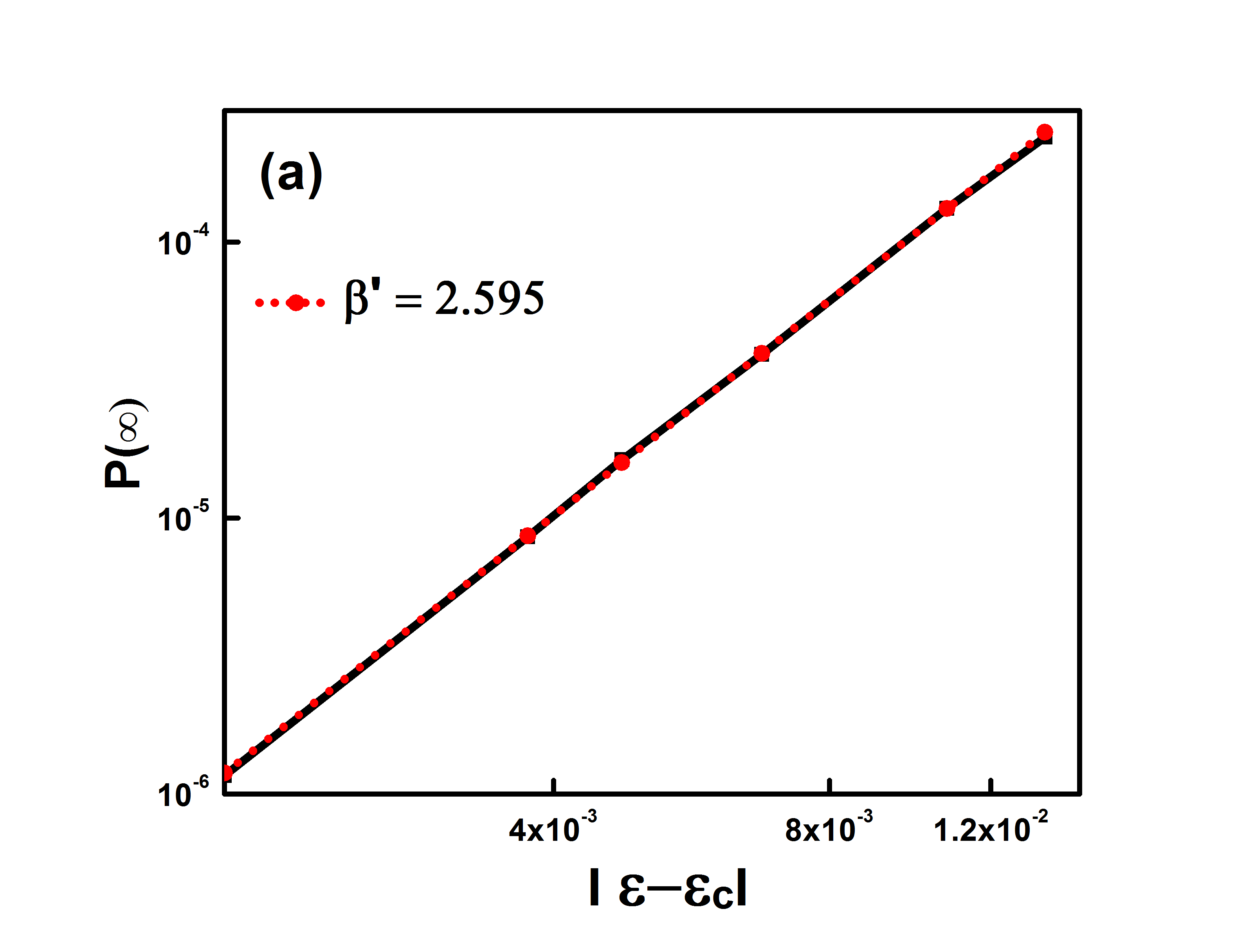}}
	\scalebox{0.235}{
		\includegraphics{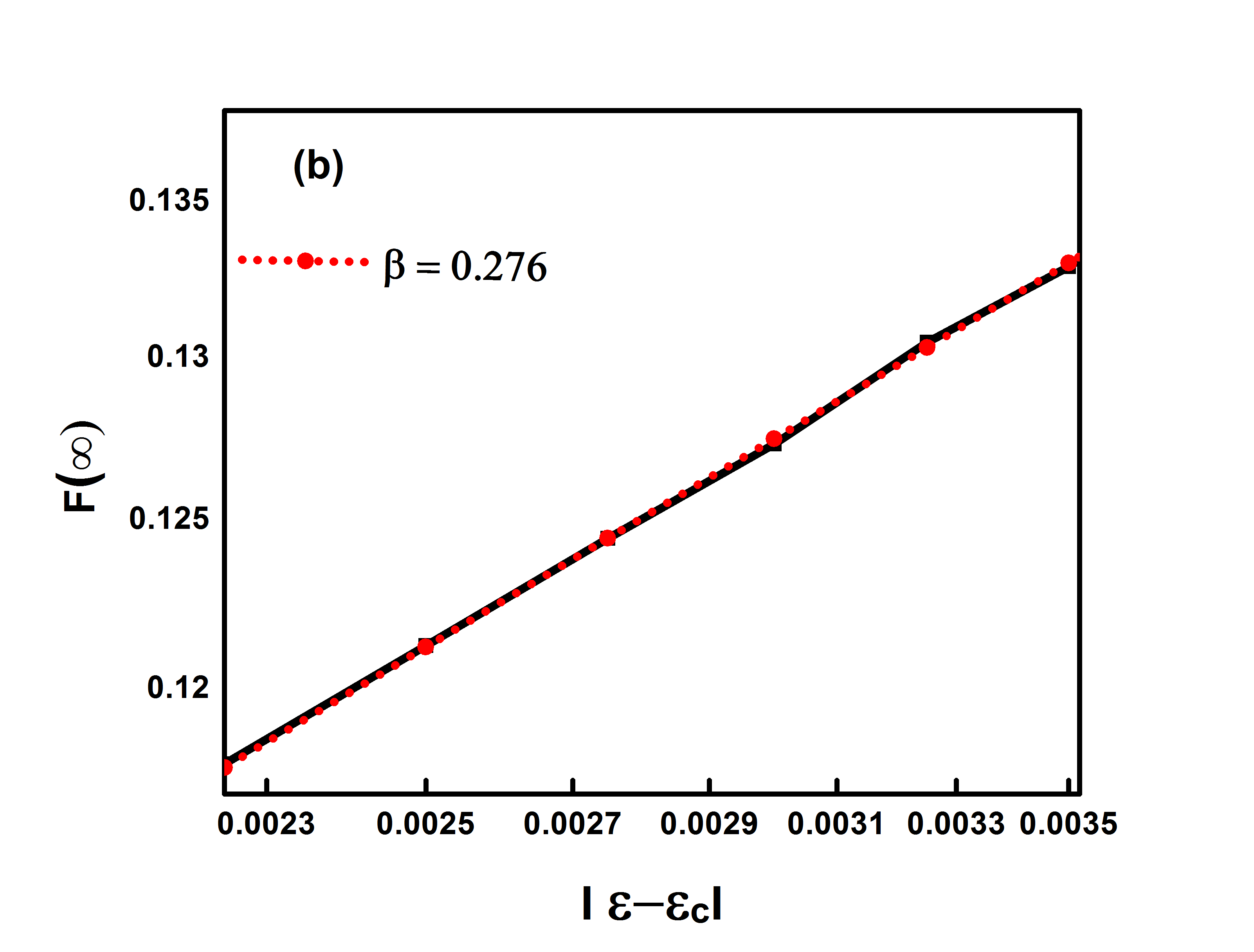}}
	\caption{ We simulate a lattice of size $N=1\times10^5$ and each 
		average $200$ configurations for several values of $\epsilon$
		above and below $\epsilon_c$.
		\textbf{(a)} We plot $\Delta$=$\lvert \epsilon-\epsilon_c \rvert$ versus $P(\infty)$. We find $\beta'=2.595$.
		\textbf{(b)} We plot $\Delta$=$\lvert \epsilon-\epsilon_c \rvert$ versus $F(\infty)$. We find $\beta=0.276$.}
	\label{fig8}
\end{figure}

We expect the flip rate $F(t)$ scale as $F(\infty) \propto \Delta^{\beta}$ , where $\Delta$=$\lvert \epsilon-\epsilon_c \rvert$ and $\beta$ = $\nu_{\parallel}$ $\delta$. Similarly, the persistence $P(\infty) \propto \Delta^{\beta'}$, where $\Delta$=$\lvert \epsilon-\epsilon_c \rvert$ and $\beta'$ = $\nu_{\parallel}$ $\theta$. We consider $N= 2\times 10^5$ and average over a $50$ configurations for both $F(t)$ and $P(t)$. We plot $\Delta$=$\lvert \epsilon-\epsilon_c \rvert$ versus $P(\infty)$ in \textbf{Fig. \ref{fig8} (a)}. The exponent $\beta'=2.595$. Similarly, we plot $\Delta$=$\lvert \epsilon-\epsilon_c \rvert$ versus $F(\infty)$ in \textbf{Fig. \ref{fig8} (b)}. The exponent $\beta$ is found to be $0.276$. These exponents are close to the expected value. They confirm the expected value of $\nu_{\parallel}$

\subsection{Coupled map lattice with 3-site interaction}

\begin{figure}[h]%
	\centering
	\includegraphics[width=0.65\textwidth]{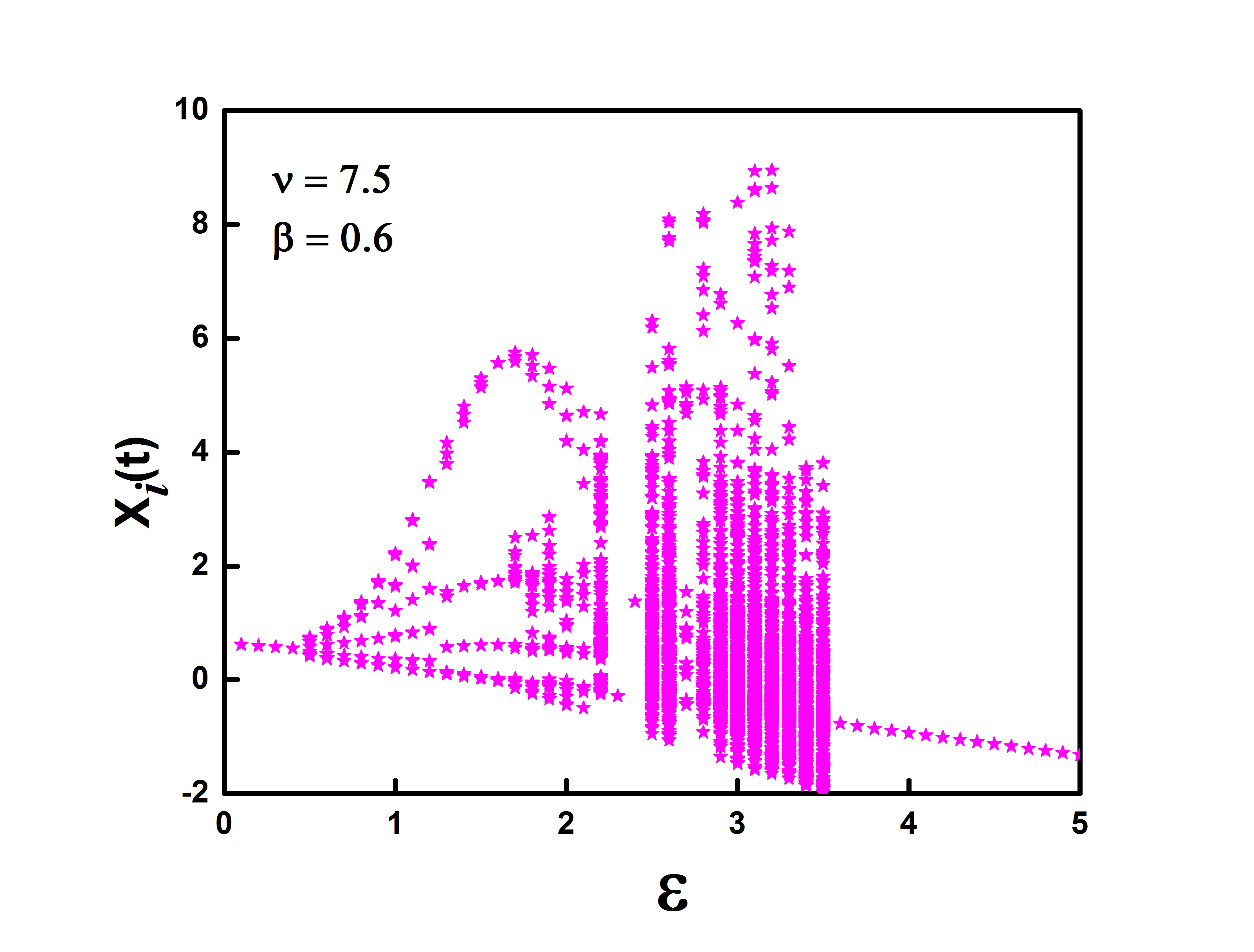}
	\caption{Bifurcation diagram for 3-site interaction coupled Gauss maps where $x_i(t)$ is plotted as a function of $\epsilon$. We fix $\nu=7.5$ and $\beta=0.6$ and simulate for $N=500$ and wait for $t=10^5$ time steps.}
	\label{fig9}
\end{figure}

We also explore coupled Gauss maps with three-site interactions. We choose the map parameters of the Gauss map $\nu=7.5$,  $\beta=0.6$, and plot the bifurcation diagram. We consider a lattice size of $N=500$ and wait for $t=10^5$ time steps. The resulting bifurcation plot displays all values of sites as a function of coupling strength $\epsilon$. We plot the bifurcation diagram in \textbf{Fig. \ref{fig9}}. We observe the series of transitions from a fully synchronized state to a chaotic state by varying a coupling parameter. For larger values of $\epsilon$, the system again displays a fully synchronized state.

\begin{figure}[hbt!]
	\centering
	\scalebox{0.23}{
		\includegraphics{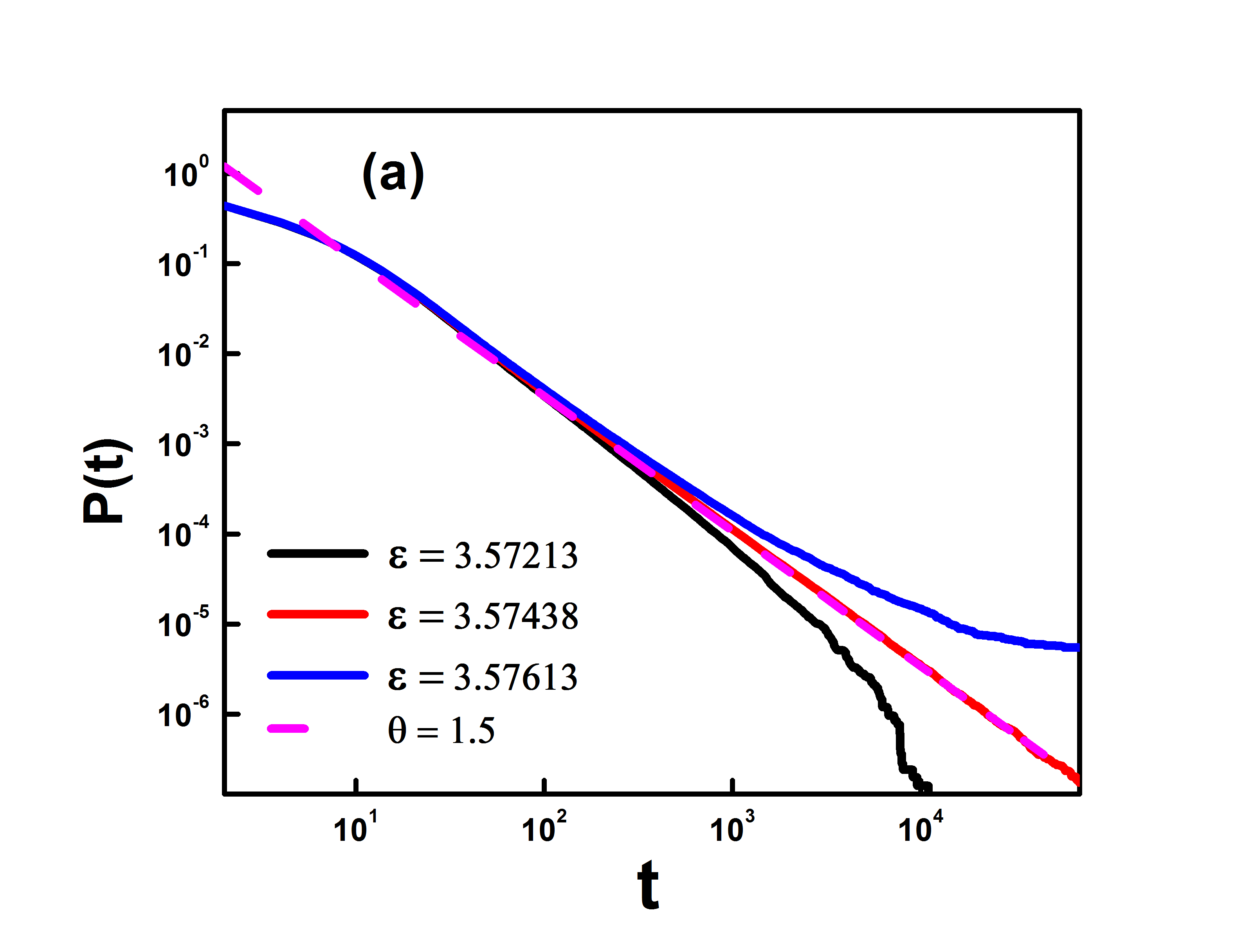}}
	\scalebox{0.23}{
		\includegraphics{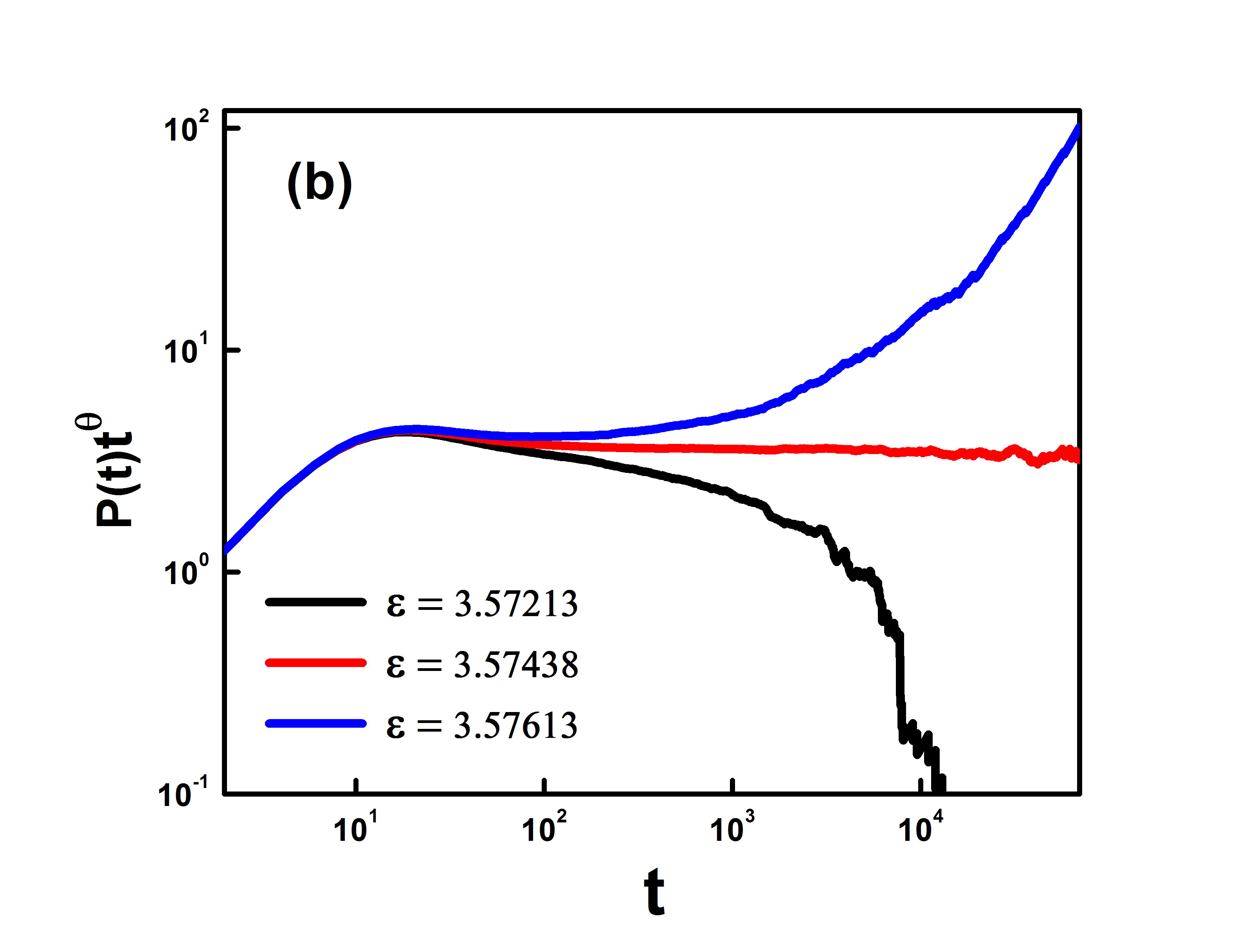}}
	\caption{ \textbf{(a)} Plot of persistence $P(t)$ as a function of time $t$ for
		$\epsilon < \epsilon_c$, $\epsilon > \epsilon_c$ and at $\epsilon=\epsilon_c$ for $N=2\times 10^5$. We average over a $400$ configuration.
		We observe that $P(t) \sim t^{-\theta}$, $\theta=1.5$ at $\epsilon=\epsilon_c=3.57438$.
		\textbf{(b)} Plot of $P(t)t^{\theta}$ as a function of $t$, for $\epsilon < \epsilon_c$,
		$\epsilon > \epsilon_c$ and at $\epsilon=\epsilon_c=3.57438$ for the same data as \textbf{(a)}.}
	\label{fig10}
\end{figure}

\begin{figure}[hbt!]
	\centering
	\scalebox{0.235}{
		\includegraphics{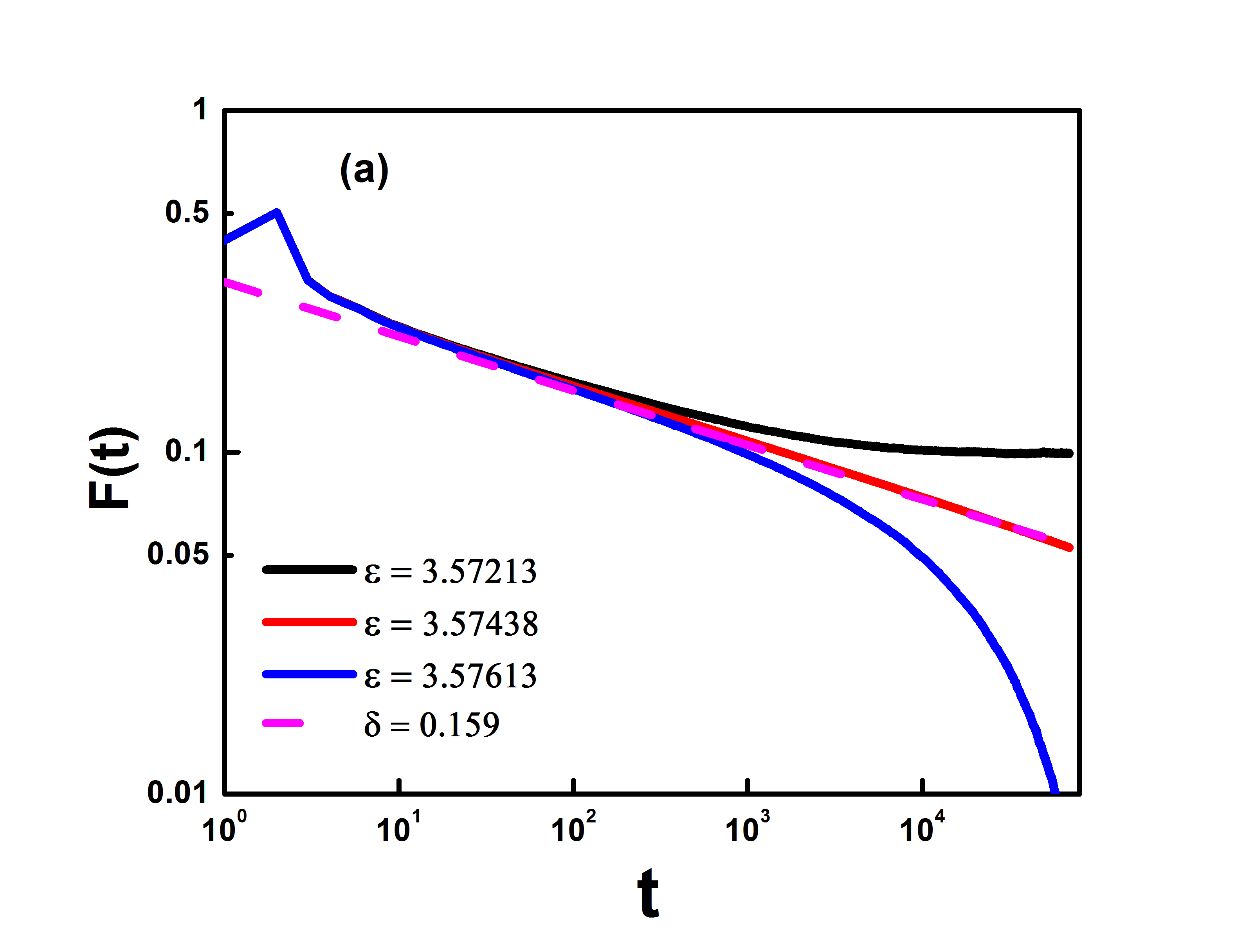}}
	\scalebox{0.235}{
		\includegraphics{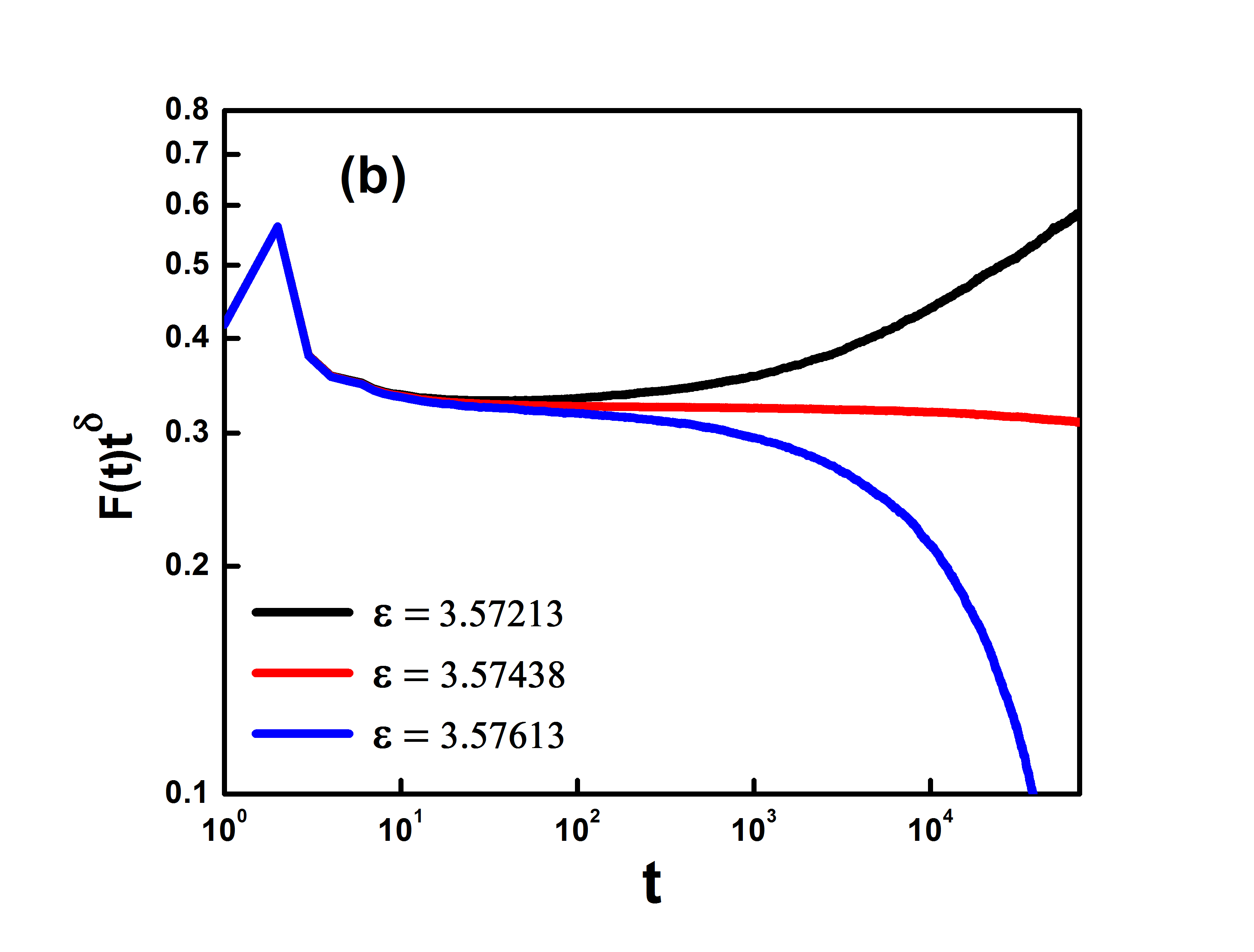}}
	\caption{ \textbf{(a)} Plot of Flip rate $F(t)$ as a function of time $t$ for
		$\epsilon < \epsilon_c$, $\epsilon > \epsilon_c$ and at $\epsilon=\epsilon_c$ for $N= 2\times 10^5$. 
		We average over a $400$ configuration.
		We observe $F(t) \sim t^{-\delta}$, $\delta=0.159$ at $\epsilon=\epsilon_c=3.57438$.
		\textbf{(b)} Plot of $F(t)t^{\delta}$ as a function of time $t$, for $\epsilon < \epsilon_c$, $\epsilon > \epsilon_c$ and at $\epsilon=\epsilon_c=3.57438$ for the same data as \textbf{(a)}}
	\label{fig11}
\end{figure}

To study the nature of phase transition for $\beta=0.6$ and $\nu=7.5$, we simulate the quantifiers $F(t)$ and $P(t)$ for distinct values of the coupling parameter $\epsilon$. We consider $N=2 \times 10^5$ and average over $400$ configurations. We plot $P(t)$ as a function of time steps $t$ in \textbf{Fig.\ref{fig10} (a)} and $F(t)$ as a function of time steps $t$ in \textbf{Fig. \ref{fig11} (a)}. From this plot, we find that, 
$F(t) \sim t^{-\delta}$ with $\delta=0.159$ and $P(t) \sim t^{-\theta}$ with $\theta=1.5$ at critical point $\epsilon=\epsilon_c=3.57438$. The obtained exponents exactly match with exponents of the DP class. We have shown $P(t)t^{\theta}$  versus $t$ and $F(t)t^{\delta}$ versus $t$ for $\epsilon < \epsilon_c$, $\epsilon > \epsilon_c$ and at $\epsilon=\epsilon_c$ in \textbf{Fig. \ref{fig10} (b)} and \textbf{Fig. \ref{fig11} (b)}. 
We observe that the quantity $F(t)$ and $P(t)$ tends to a constant
for $\epsilon=\epsilon_c$ as $t\rightarrow \infty$. For $\epsilon\neq \epsilon_c$, the curve displays upward or downward curvature.
This is alternative evidence
for the confirmation of the obtained values of the exponent.
The exponent $\delta$ and $\theta$ match with DP exactly. The multi-site 
interactions can also be termed as simplicial couplings.

\begin{figure}[hbt!]
	\centering
	\scalebox{0.235}{
		\includegraphics{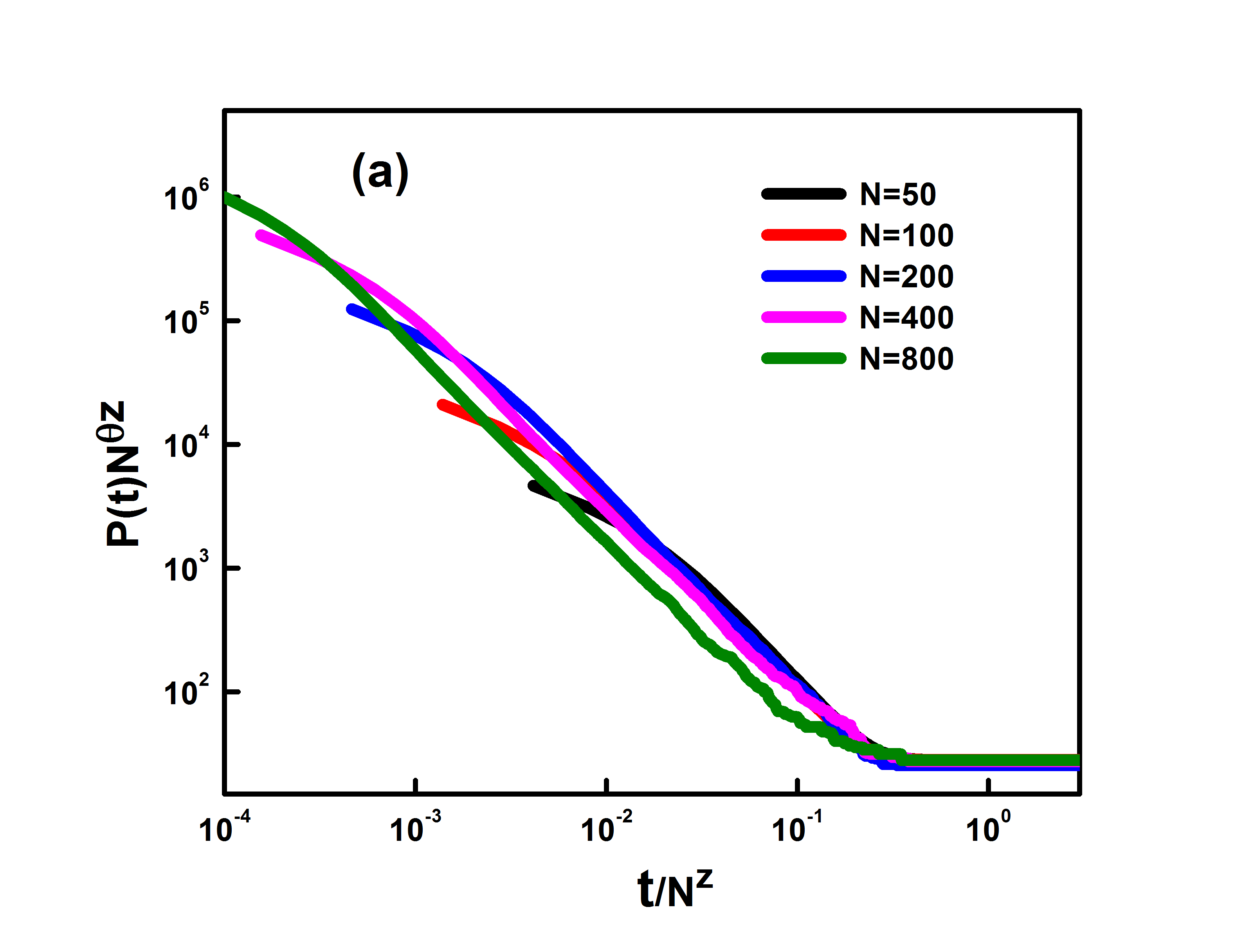}}
	\scalebox{0.235}{
		\includegraphics{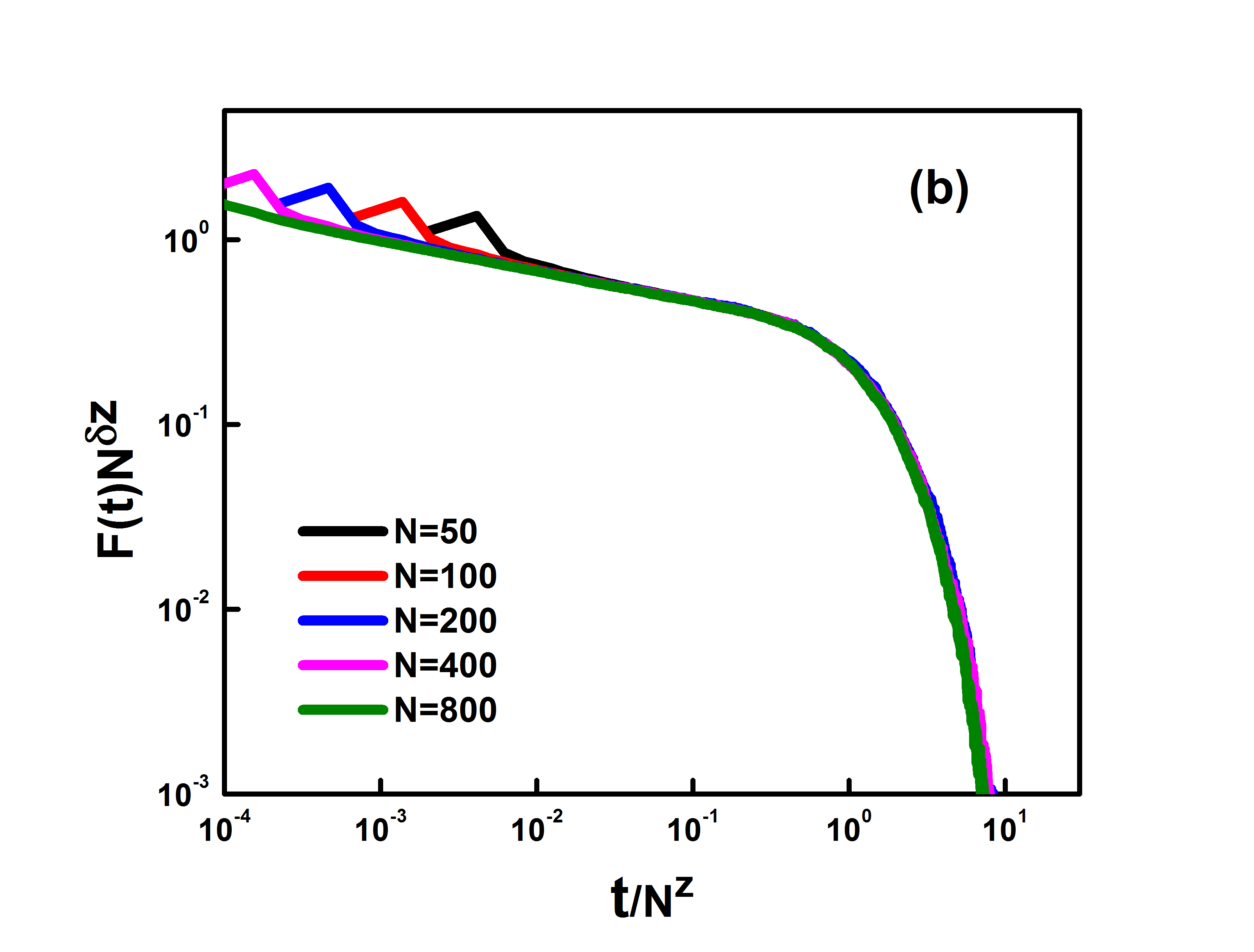}}
	\caption{\textbf{(a)} Plot of $P(t) N^{\theta z}$ as a function of
		$t/N^z$ at $\epsilon=\epsilon_c=3.57438$ and $\theta$=1.5. 
		A good scaling collapse
		is observed at $z=1.58$ and $\theta=1.5$ .
		\textbf{(b)} Plot of $F(t) N^{\delta z}$ as a function of $t/N^z$ at $\epsilon=\epsilon_c=3.57438$.
		The fine collapse is observed at $z=1.58$ and $\delta$=0.159.}
	\label{fig12}
\end{figure}

Further, we study the finite-size scaling to compute the dynamic exponent $z$. We consider $N=50,100,200,400,800$ lattice sizes and average over more than $4\times 10^4$ configurations for both $P(t)$ and $F(t)$.  We plot $P(t) N^{\theta z}$ versus $t/N^z$ in \textbf{Fig. \ref{fig12} (a)} and $F(t) N^{\delta z}$ as a function of $t/N^z$ in \textbf{Fig. \ref{fig12} (b)} at $\epsilon=\epsilon_c=3.57438$. From this plot, we observe that for $P(t)$ a good scaling collapse is observed for $z=1.58$ and $\theta=1.5$. While for $F(t)$ a good scaling collapse is observed for $z=1.58$ and $\delta=0.159$. The exponent obtained from finite size scaling for both $P(t)$ and $F(t)$ is exactly the same as the exponent of the DP universality class.

\begin{figure}[hbt!]
	\centering
	\scalebox{0.235}{
		\includegraphics{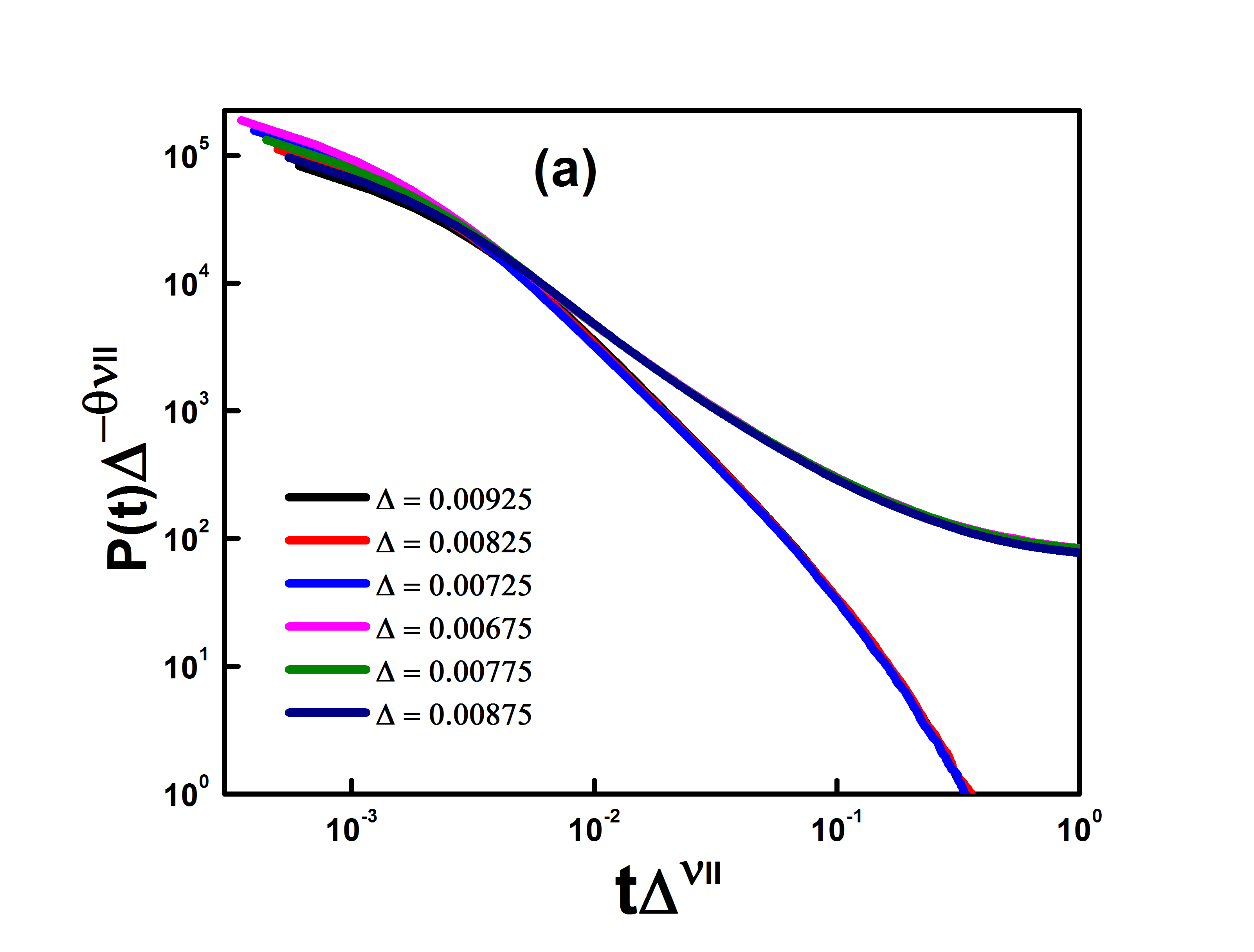}}
	\scalebox{0.235}{
		\includegraphics{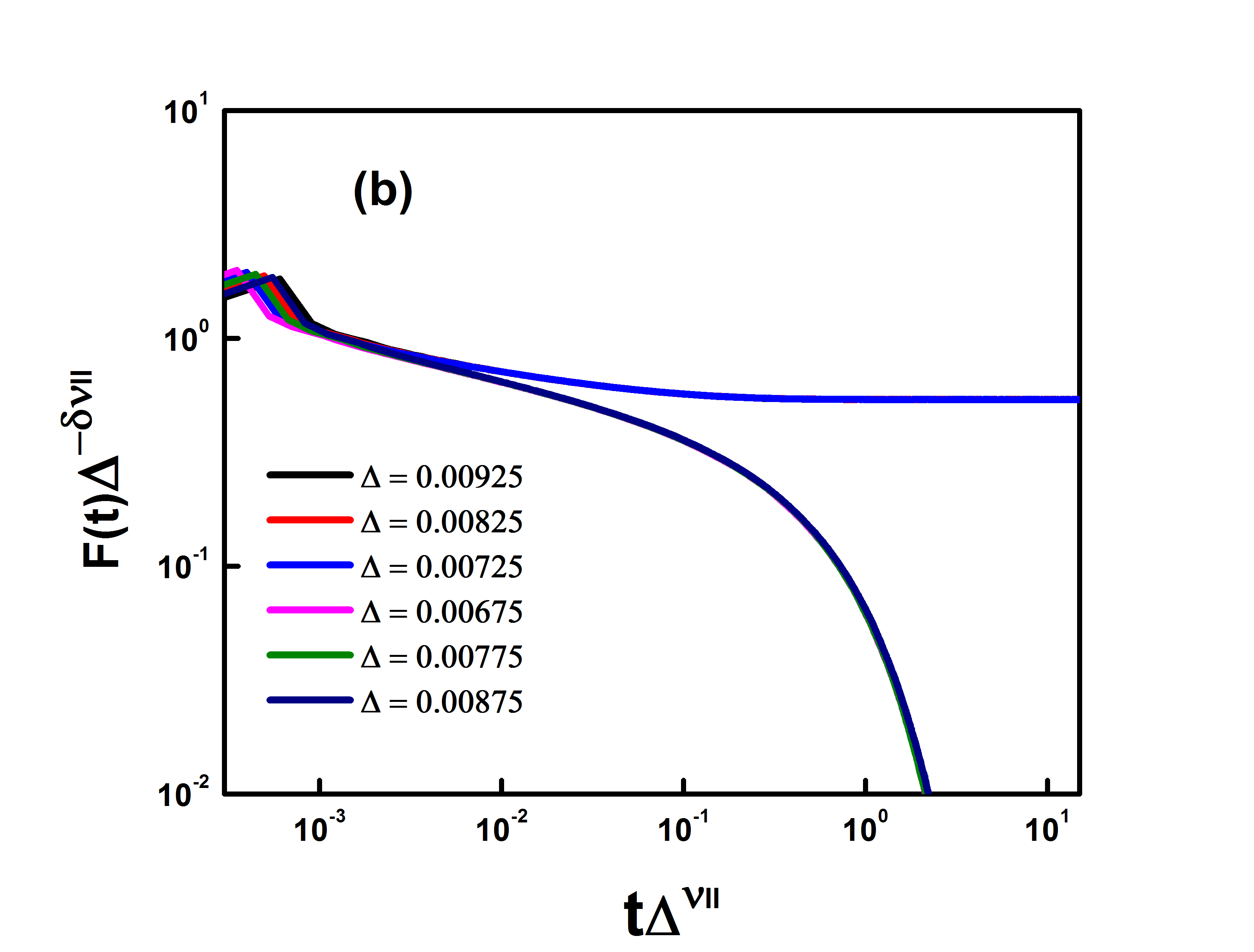}}
	\caption{ We simulate a lattice of size $N=2\times10^5$ and each 
		average over more than
		$100$ configurations for several values of $\epsilon$
		above and below $\epsilon_c$.
		\textbf{(a)}  We plot $P(t){\Delta}^{-{\theta}{\nu_{\parallel}}}$
		as a function of
		$t {\Delta}^{\nu_{\parallel}}$ for $\epsilon < \epsilon_c$
		and for	$\epsilon > \epsilon_c$.
		The good scaling collapse is observed at $\nu_{\parallel}=1.73$.
		\textbf{(b)} We plot $F(t) {\Delta}^{-{\delta}{\nu_{\parallel}}}$
		as a function of
		$t {\Delta}^{\nu_{\parallel}}$.
		The good scaling collapse is observed at $\nu_{\parallel}=1.73$.}
	\label{fig13}
\end{figure}

We study off-critical scaling to compute the exponent $\nu_{\parallel}$. 
We simulate the lattice for distinct values of $\epsilon$ for above and below the critical point. We consider $N=2\times10^5$ and average over a $100$ configuration.
We plot $P(t){\Delta}^{-{\theta}{\nu_{\parallel}}}$
as a function of $t {\Delta}^{\nu_{\parallel}}$
and the good scaling collapse is obtained
for $\nu_{\parallel}=1.73$. This behavior is shown in \textbf{ Fig. \ref{fig13} (a)}.
Similarly, we plot $F(t){\Delta}^{-{\delta}{\nu_{\parallel}}}$
as a function of
$t {\Delta}^{\nu_{\parallel}}$ and a good scaling collapse is
obtained at $\nu_{\parallel}=1.73$.
This collapse is shown in \textbf{Fig. \ref{fig13} (b)}.
The obtained values of $\nu_{\parallel}=1.73$ match with
the exponent of the DP class exactly. 

\begin{figure}[hbt!]
	\centering
	\scalebox{0.235}{
		\includegraphics{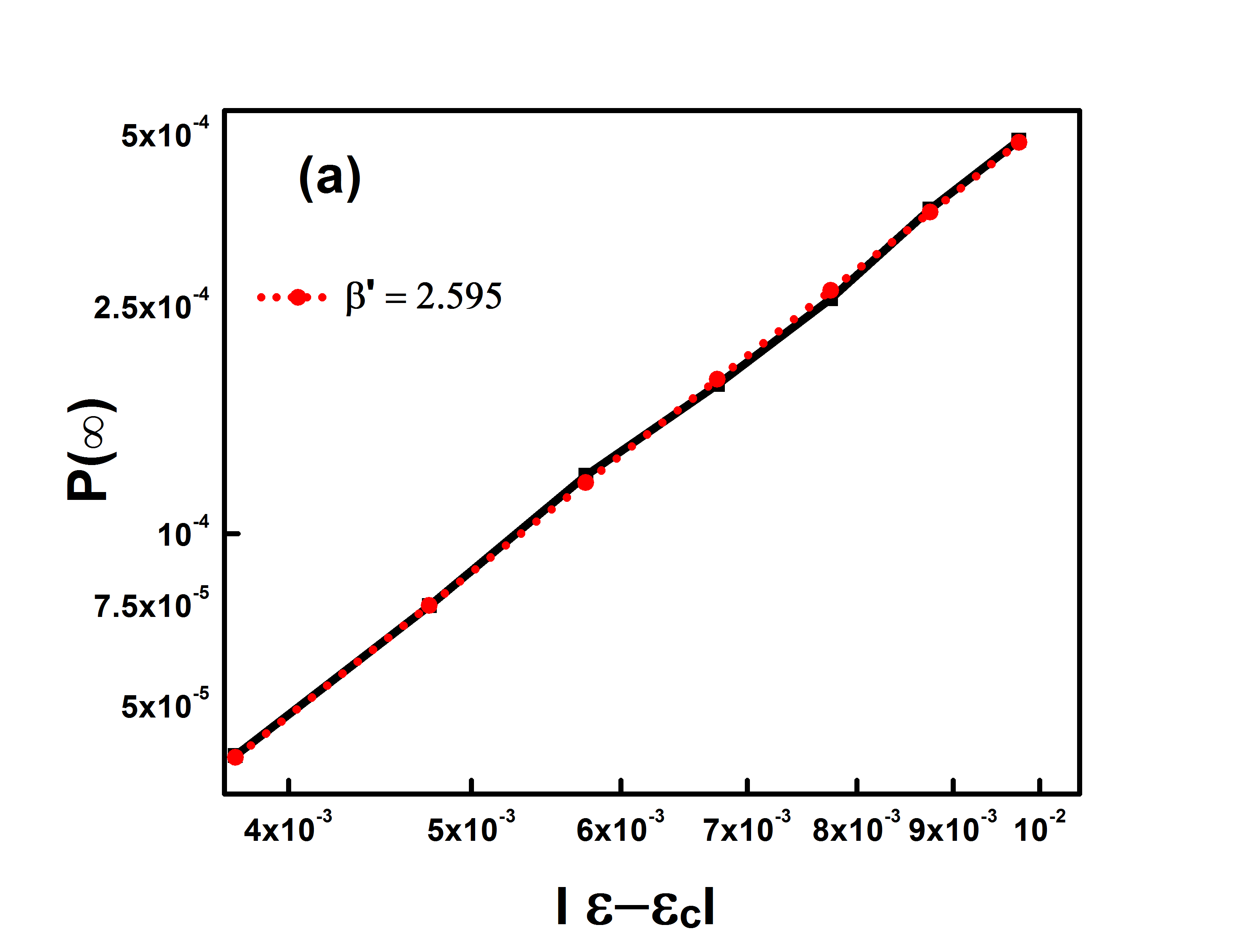}}
	\scalebox{0.235}{
		\includegraphics{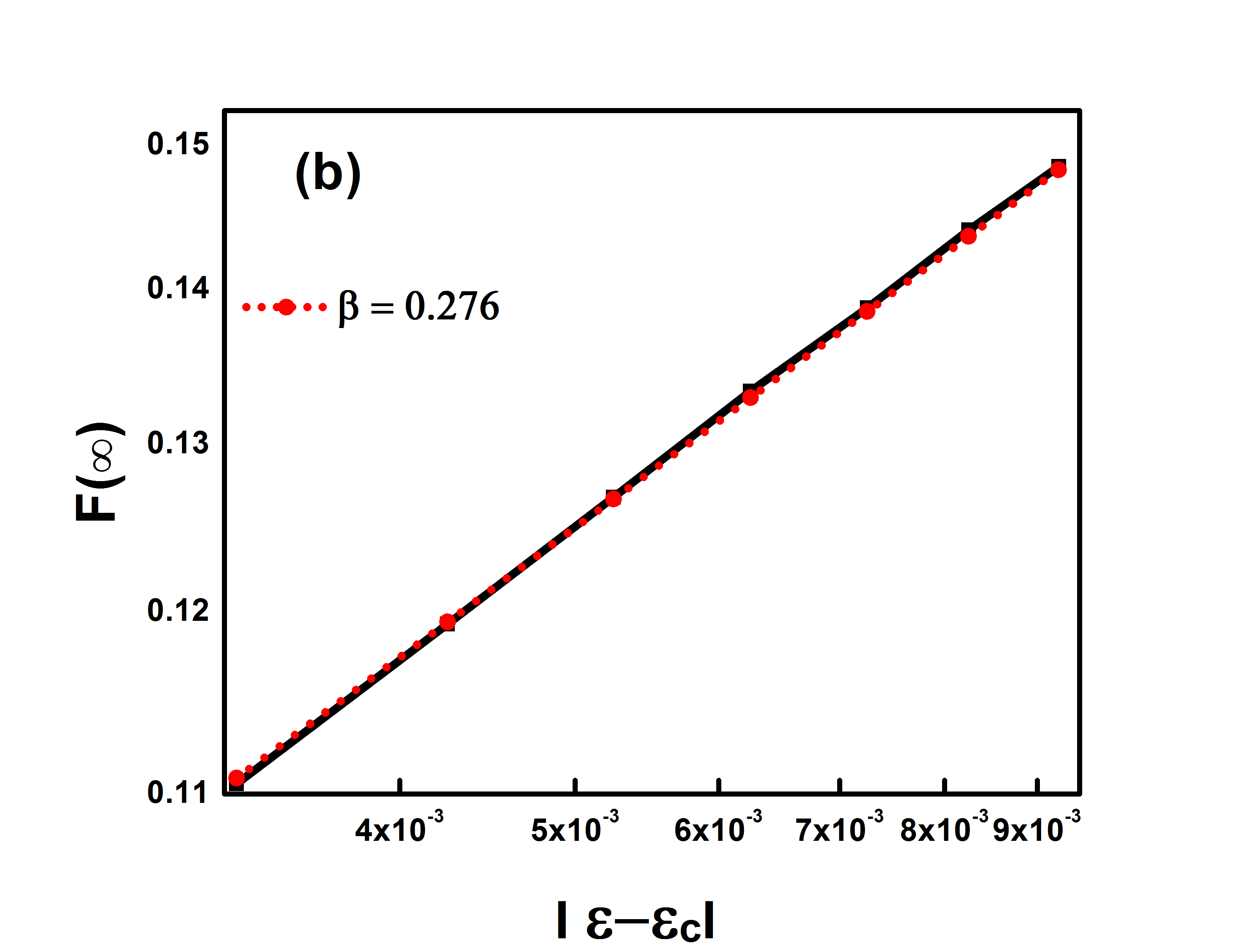}}
	\caption{ We simulate a lattice of size $N=1\times10^5$ and each 
		average $200$ configurations for several values of $\epsilon$
		above and below $\epsilon_c$.
		\textbf{(a)} We plot $P(\infty)$ versus $\Delta$=$\lvert \epsilon-\epsilon_c \rvert$. We find $\beta'=2.595$.
		\textbf{(b)} We plot $F(\infty)$ versus $\Delta$=$\lvert \epsilon-\epsilon_c \rvert$. We find $\beta=0.276$.}
	\label{fig14}
\end{figure}

We expect the flip rate $F(t)$ scale as $F(\infty) \propto \Delta^{\beta}$ , where $\Delta$=$\lvert \epsilon-\epsilon_c \rvert$ and $\beta$ = $\nu_{\parallel}$ $\delta$. Similarly, the persistence $P(\infty) \propto \Delta^{\beta'}$, where $\Delta$=$\lvert \epsilon-\epsilon_c \rvert$ and $\beta'$ = $\nu_{\parallel}$ $\theta$. We consider $N= 2\times 10^5$ and average over a $50$ configurations for both $F(t)$ and $P(t)$. We plot $P(\infty)$ versus $\Delta$=$\lvert \epsilon-\epsilon_c \rvert$  in  \textbf{ Fig. \ref{fig14} (a)} and the exponent $\beta'$ is found to be $2.595$. Similarly, we plot $F(\infty)$ versus $\Delta$=$\lvert \epsilon-\epsilon_c \rvert$ in \textbf{ Fig. \ref{fig14} (b)} and the exponent $\beta$ is found to be $0.276$.  This is further
confirmation of the exponent $\nu{\parallel}$

\subsection{Coupled map lattice with 4-site interaction}
\begin{figure}[h]%
	\centering
	\includegraphics[width=0.65\textwidth]{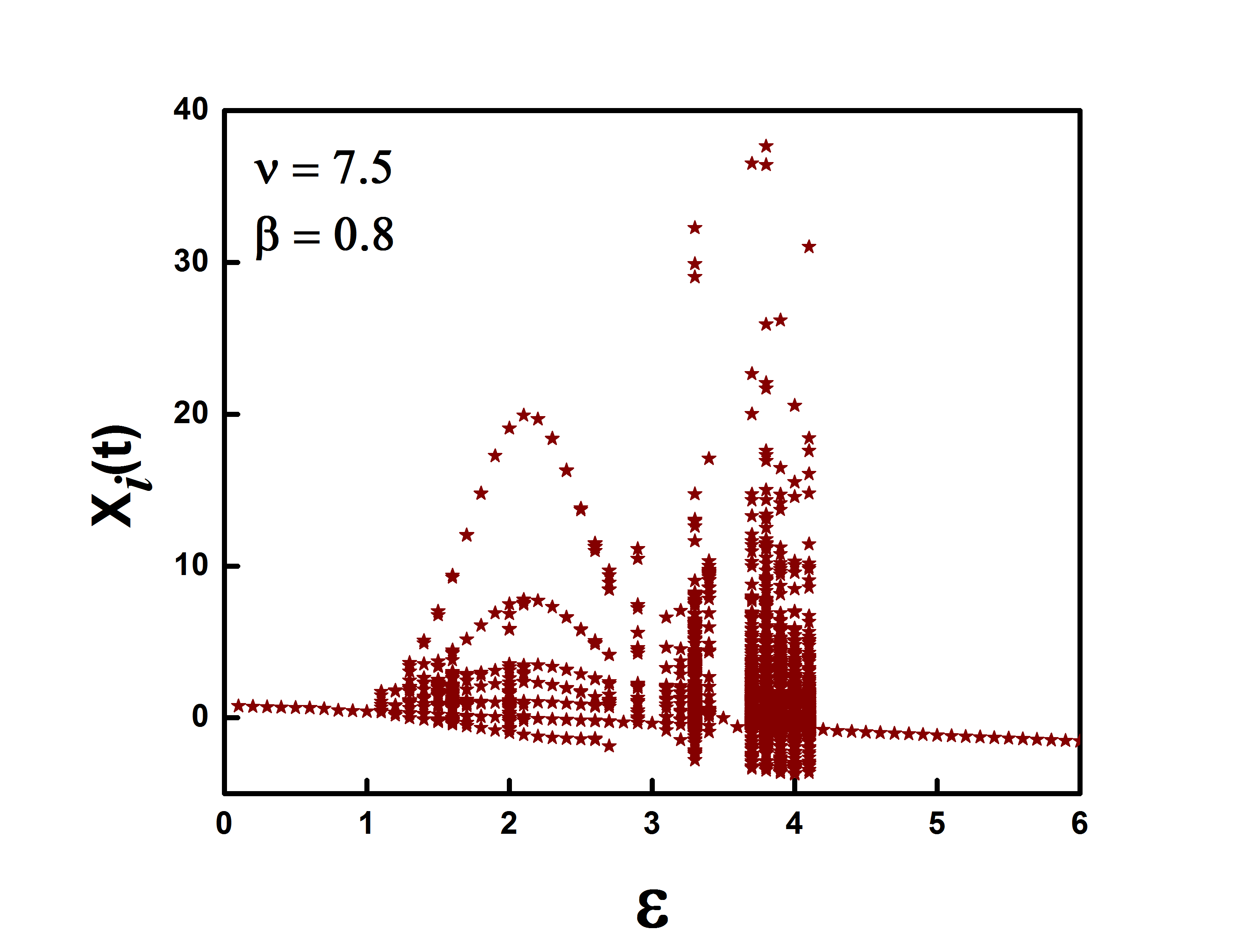}
	\caption{Bifurcation diagram for 4-site interaction coupled Gauss maps where $x_i(t)$ is plotted as a function of $\epsilon$. We fix $\nu=7.5$ and $\beta=0.8$ and simulate for $N=500$ and wait for $t=10^5$ time steps.}
	\label{fig15}
\end{figure}

We also investigate the nature of the phase transition for even site interaction in a coupled Gauss map. We simulate our model for 4-site interaction.
We choose the map parameters $\beta=0.8$ and $\nu=7.5$ and plot the bifurcation diagram for $N=500$ after waiting for $t=10^5$ steps. We plot the bifurcation diagram in \textbf{Fig. \ref{fig15}}. The bifurcation diagram is useful for visualization of the behavior of the system as a function of the coupling parameter. The bifurcation diagram captures the transition from a fully synchronized state to chaos. As the coupling parameter is increased further, the system reaches a synchronized state. We focus on one of the transitions to synchronization that shows continuous transition. 

\begin{figure}[hbt!]
	\centering
	\scalebox{0.235}{
		\includegraphics{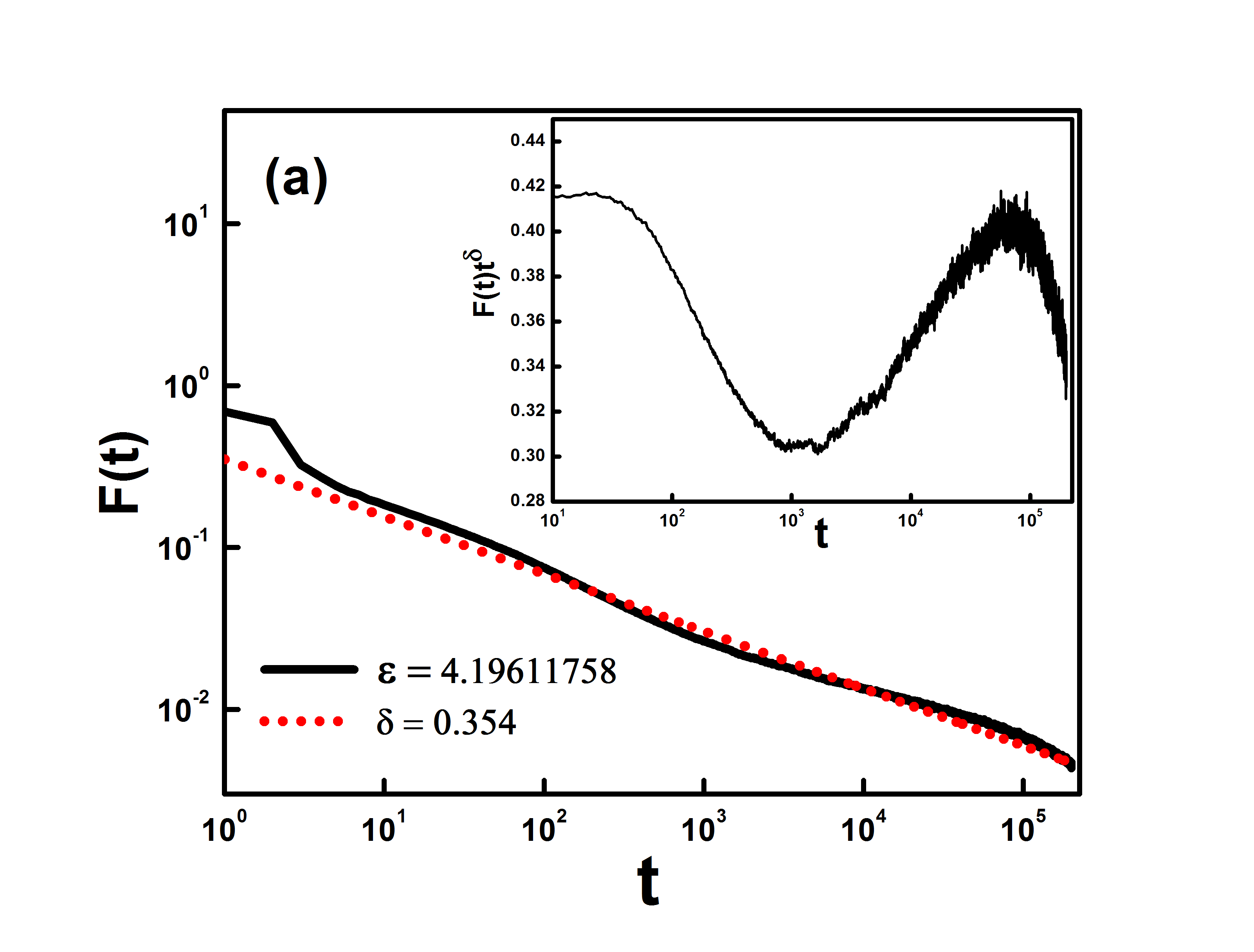}}
	\scalebox{0.235}{
		\includegraphics{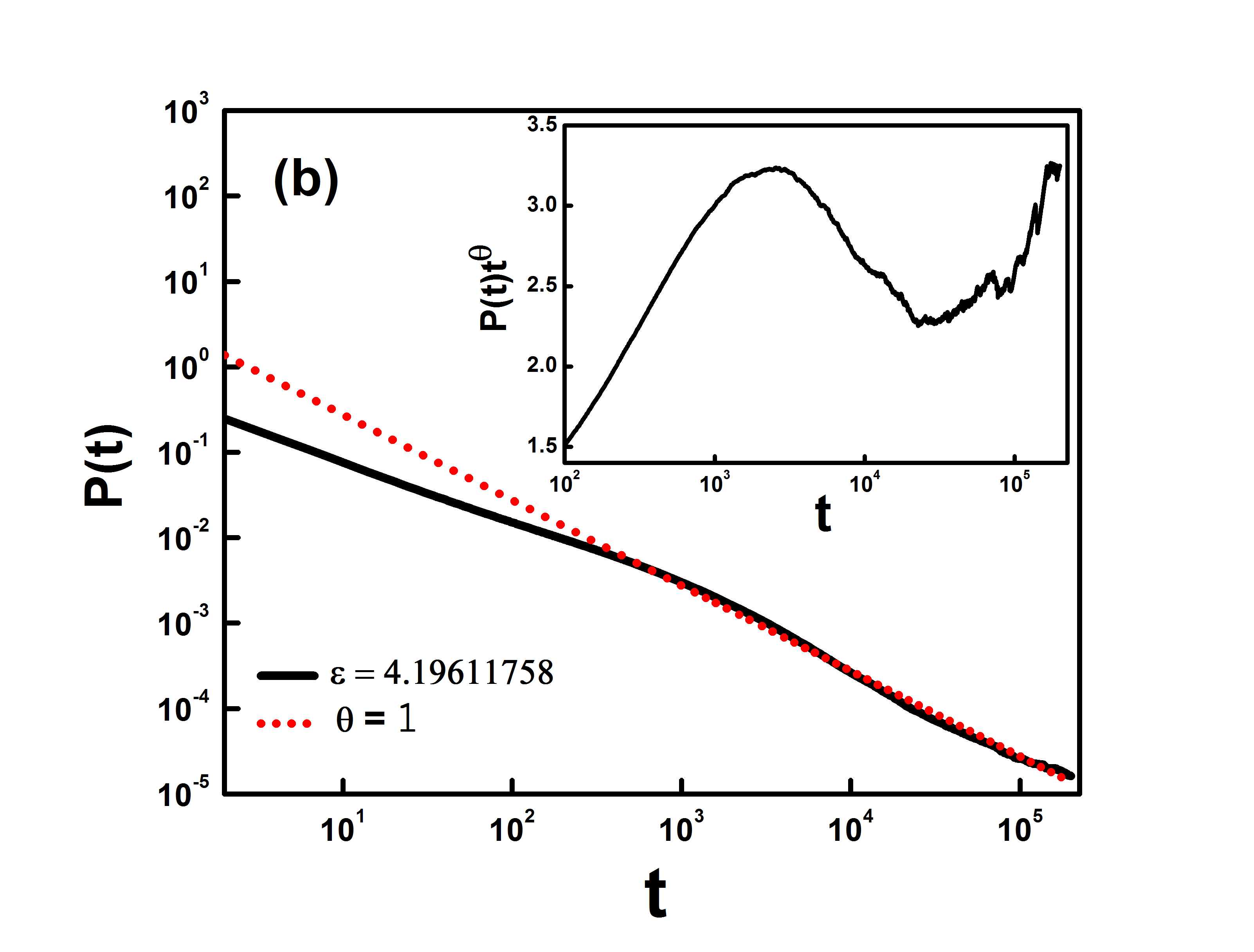}}
	\caption{Plot of 4-site interaction coupled Gauss map for $N= 2\times 10 ^5$ and average over 200 configurations. \textbf{(a)} Main Figure: Plot of $F(t)$ as a function of $t$ at $\epsilon=\epsilon_c=4.19611758$. $F(t)$ show the clear logarithmic oscillation with $\delta=0.354$. Inset: Plot of $F(t)t^{\delta}$ as a function of $t$. \textbf{(b)} Main Figure: Plot of $P(t)$ as a function of $t$, at $\epsilon=\epsilon_c=4.19611758$. $P(t)$ show the clear logarithmic oscillation with $\theta=1$. Inset: Plot of $P(t)t^{\theta}$ as a function of $t$.}
	\label{fig16}
\end{figure}

We compute the flip rate $F(t)$ and persistence $P(t)$ as the order parameter at the transition point for this system. We consider $N=2 \times10^5$ and average over a $200$ configuration. We plot $F(t)$ as a function of time steps $t$ in \textbf{Fig. \ref{fig16} (a)}. We observe that $F(t)$ shows power-law decay with logarithmic oscillation at $\epsilon=\epsilon_c=4.19611758$. The decay exponent is found to be $\delta=0.354$. We have displayed the $F(t)t^{\delta}$ as a function of $t$ for the same data in the inset plot. This plot confirms a clear oscillation over and above the power-law. Similarly, we plot $P(t)$ as a function of time $t$ in \textbf{Fig. \ref{fig16} (b)}. We find that $P(t)$ shows power-law decay superposed with logarithmic oscillations at $\epsilon=\epsilon_c=4.19611758$. The decay exponent is found to be $\theta=1$. We also plot $F(t)t^{\theta}$ as a function of $t$ for the same data in the inset plot. This plot displays oscillations over and above the power-law that confirm the $\theta$ values and logarithmic oscillation. This provides a novel transition with an exponent that does not match DP for both 
persistence exponent $\delta$ or $\theta$. Therefore, the 4-site interaction is a relevant perturbation that emerges in a novel transition for the Gauss map.
 
\subsection{Coupled map lattice with 2-site interaction}
\begin{figure}[h]%
	\centering
	\includegraphics[width=0.65\textwidth]{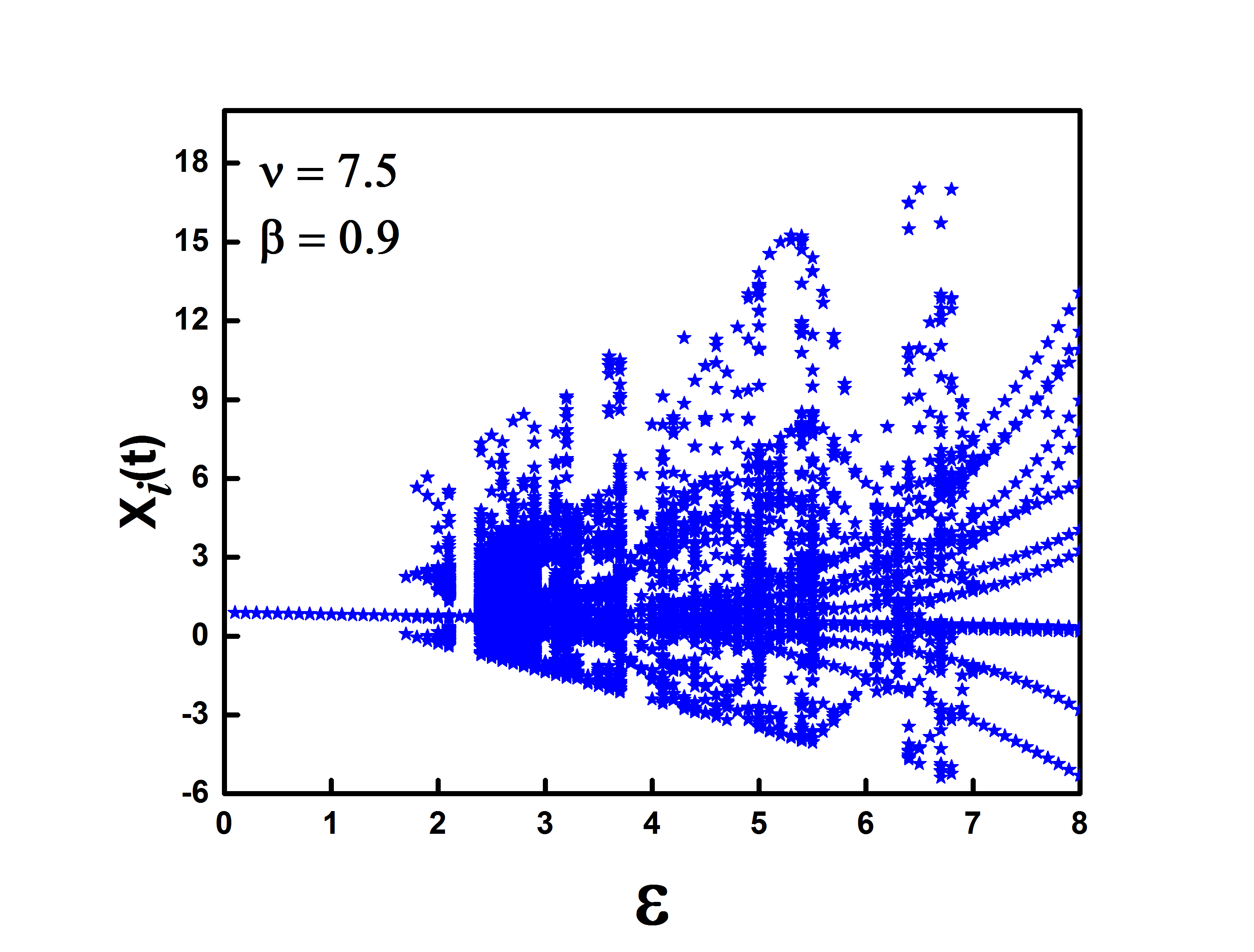}
	\caption{Plot of bifurcation diagram for 2-site interaction coupled Gauss maps where $x_i(t)$ is plotted as a function of $\epsilon$. We fix $\nu=7.5$, $\beta=0.9$ and simulate for $N=500$ and $t=10^5$ time steps.}
	\label{fig17}
\end{figure}

Lastly, we investigate the model with 2-site interaction. We consider $N=500$, wait for $t=10^5$, and plot the bifurcation diagram for $\beta=0.9$ and $\nu=7.8$. The bifurcation diagram depicts the behavior of the system as a function of the coupling strength. \textbf{Fig. \ref{fig17}} shows the bifurcation diagram. From this plot, we find that the system displays a fully synchronized state for low coupling. A transition occurs at $\epsilon$ = 1.5 and again for 2.1 $\le$ $\epsilon$ $\le$ 2.4, and the system transitions to a synchronized state. We again investigate the point that shows a continuous transition.
 
\begin{figure}[hbt!]
	\centering
	\scalebox{0.235}{
		\includegraphics{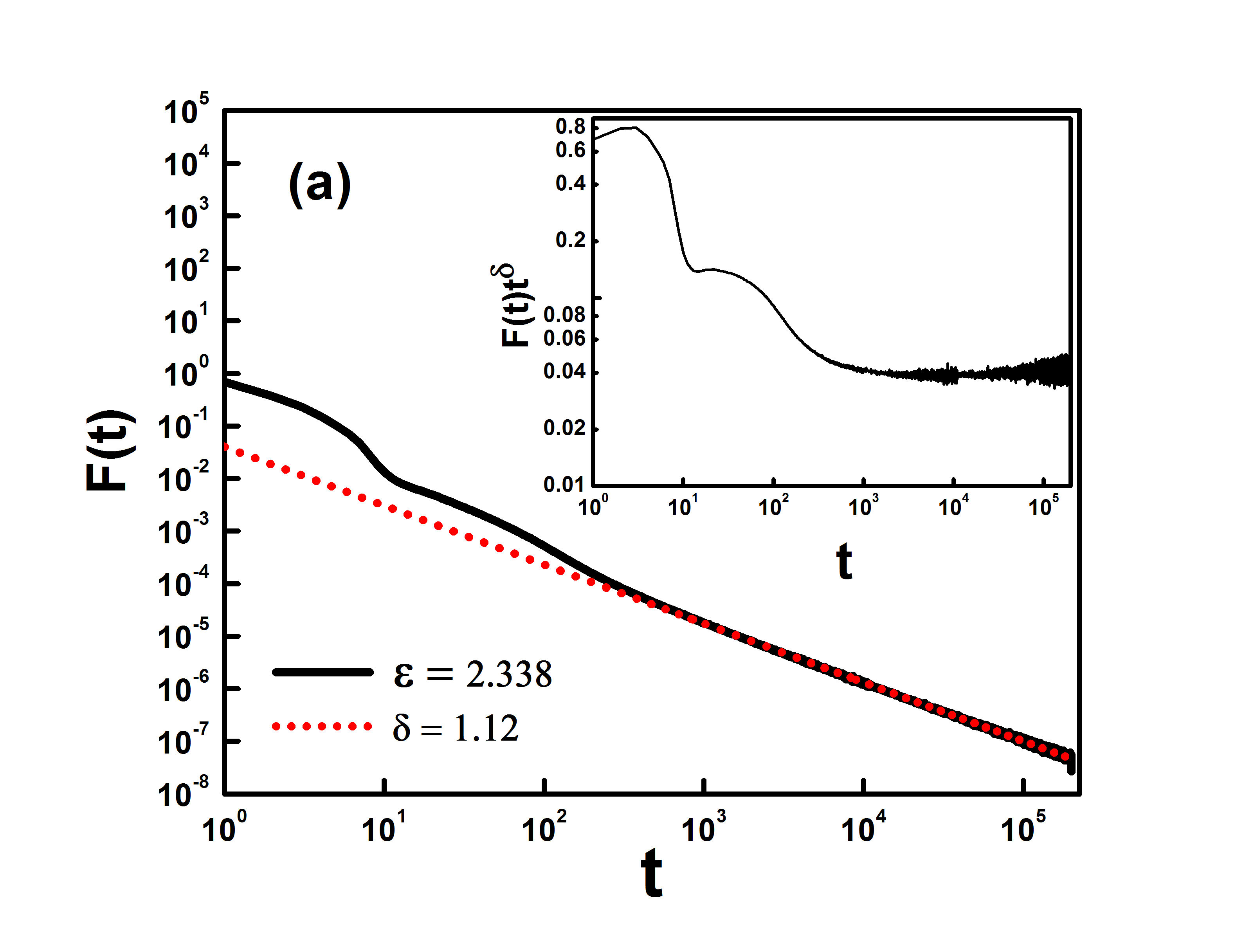}}
    \scalebox{0.235}{
		\includegraphics{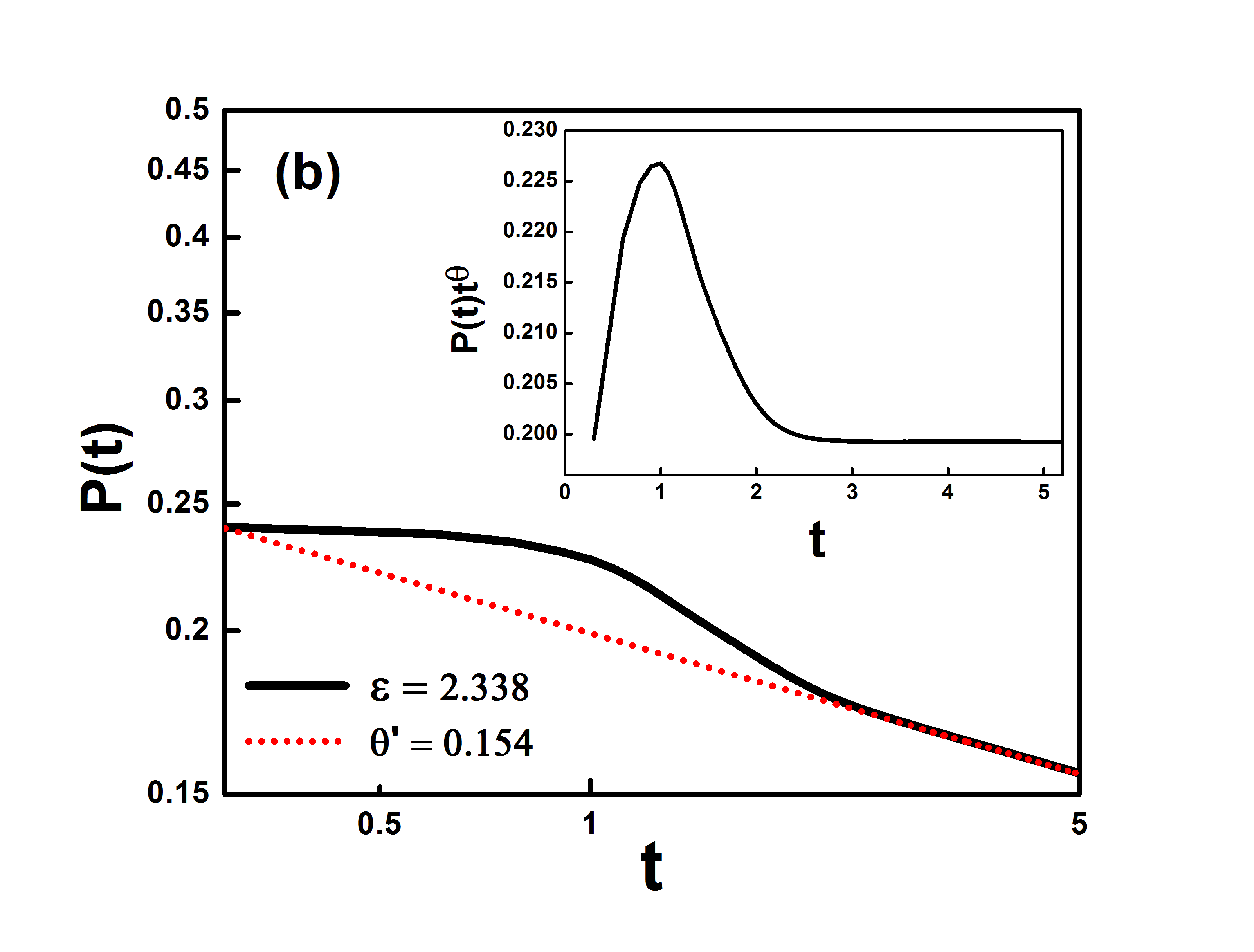}}
	\caption{Plot of 2-site interaction coupled Gauss map for $N=4 \times10^4$ and average over a $1200$ configuration. \textbf{(a)} Main Figure: Plot of $F(t)$ as a function of $t$ for $\epsilon=\epsilon_c=2.338$. We carry out a running time average to reduce fluctuations. We find $F(t) \sim t^{-\delta}$  with $\delta=1.12$. Inset: $F(t)t^{\delta}$ as a function of $t$. 
    \textbf{(b)} Main Figure: Plot of $P(t)$ as a function of $\log(t)$ for $\epsilon=\epsilon_c=2.338$. We find $P(t) \sim 1/\log (t)^{\theta'}$  with $\theta'=0.154$. Inset: Plot of $P(t) \log(t)^{\theta'}$ as a function of $t$.}
	\label{fig18}
\end{figure}

We compute the flip rate $F(t)$ and persistence $P(t)$ for $N=4 \times10^4$ and average over a $1200$ configuration. \textbf{Fig. \ref{fig18} (a) } shows the plot $F(t)$ as a function of time $t$. The inset on the main figure shows the plot of $F(t)t^{\delta}$ as a function of $t$.  We find $F(t) \sim t^{-\delta}$ at $\epsilon=\epsilon_c=2.338$. The decay exponent is found to be 
$\delta=1.12$. This exponent is very different from the DP class. 
We also plot $P(t)$ as a function of $\log(t)$ in \textbf{Fig. \ref{fig18} (b)}. It decays as $\log(t)^{-\theta'}$ The inset  shows a plot of $P(t)\log(t^{\theta'})$ as a function of $t$ with $\theta'=0.154$. The persistence decays slower than power-law in this case and there is no well-defined persistence exponent.

\section*{Summary}

Directed percolation is one of the most widely observed non-equilibrium dynamic transitions to an absorbing state. Of late,
it has been observed in experimental situations as well. 
Most of these systems are modeled by equations involving a Laplacian. 
The standard formulation of coupled map lattices can be considered as 
discretized Lapcian. It involves pairwise 
interactions. This model is not a discretized Laplacian and the interactions are multi-body multiplicative interactions. 

In this work, we study the coupled Gauss map for odd and even site interaction in one dimension. We study 3-site and 5-site for odd sites and 2-site and 4-site for even sites interaction. The coupling is such that it cannot be decomposed into pairwise interactions. We have classified the sites, considering fixed points as a reference. If the variable values are above or below the fixed point, then the associate spin is $(+1)$ or $(-1)$. We investigate the phase transition using quantifiers such as the flip rate $F(t)$ and persistence $P(t)$. 

\textbf{$i)$} For 3-site and 5-site interaction, we find the power-law decay of flip rate $F(t)$ and persistence $P(t)$ at a critical point. The decay exponent for $F(t)$ is $\delta$=0.159 and for $P(t)$ is $\theta$=1.5. We also study the finite-size scaling and the off-critical scaling. The obtained exponents $z=1.5$, $\nu_{\parallel}=1.73$, and $\beta$=0.275 match exactly with the exponents of the DP universality class. Thus DP class is robust for 3-site and 5-site interaction studied in this work. 

\textbf{$ii)$} For even site interaction, the exponents change. For 4-site interaction, we find the decay exponent $\delta=0.354$ and $\theta=1$. We observe the logarithmic oscillations over and above the power-law. For 2-site interaction, we find the decay exponent $\delta$=1.12. The persistence decays slower than power-law as
$\log(t)^{-\theta'}$. The behaviors are different from DP and they are different from each other. Thus, multi-site couplings could lead to a new universality class. It could depend on the number of coupled sites. It could also depend on the detailed nature of the map. Further studies on different maps as well as interactions could be of interest. The results are quite surprising given the fact that only a few 
universality classes are known for absorbing state transitions.

\section*{Acknowledgments}
PMG thanks DST-SERB (CRG/2020/003993) for financial assistance. 

\section*{Author Contribution}
MCW and PMG contributed
to conceptualization, simulations, visualization, and drafting.

\section*{Statements and Declarations}
The authors declare that they have no conflict of interest.

\section*{Data Availability Statement:} This manuscript has no associated data or
the data will not be deposited.
[Author's comment: All data that support the plots within the paper and 
other findings of the study are available from the first author upon reasonable
request.]
\bibliographystyle{unsrt}
\bibliography{gaussref}

\end{document}